\documentclass[aps,amsmath,prb,a4paper,floatfix,amssymb,twocolumn]{revtex4}
\usepackage{graphicx}
\usepackage{subfigure}
\usepackage{stackengine}
\usepackage{amsmath}
\usepackage{amsfonts}
\usepackage{amssymb}
\usepackage{xcolor}
\newcommand{\beq}{\begin{equation}}
\newcommand{\eneq}{\end{equation}}
\newcommand{\bea}{\begin{eqnarray}}
\newcommand{\enea}{\end{eqnarray}}
\newcommand{\met}{\frac{1}{2}} 
\newcommand{\freccia}{ \quad  \quad \mathbf{\Rightarrow} \quad }

\newcommand{\ppsi}{{\Psi '}}
\newcommand{\pphi}{{\Phi '}}
\newcommand{\kp}{{k_\parallel}}

\newcommand{\ax}{\`a~}

\UseRawInputEncoding

\begin{document}
\title{ Quantum zero point  electromagnetic  energy  difference  between the superconducting and the normal phase in  a   HTc superconducting  metal bulk sample}
 \author{ {Annalisa}  {Allocca,}$^{1,2}$
{Saverio}  {Avino,}$^{2,3}$
{Sergio}  {Balestrieri,}$^{1,4}$
{Enrico}  {Calloni,}$^{1,2}$
{Sergio}  {Caprara,}$^{5,6}$
{Massimo}  {Carpinelli,}$^{7,8}$
 {Luca}  {D'Onofrio,}$^{1,2}$
{Domenico}  {D'Urso,}$^{8,9}$
{Rosario}  {De Rosa,}$^{1,2}$
{Luciano}  {Errico,}$^{1,2}$
{Gianluca}  {Gagliardi,}$^{2,3}$
{Marco}  {Grilli,}$^{5,6}$
{Valentina}  {Mangano,}$^{5,6}$
{Maria}  {Marsella,}$^{5,6}$
{Luca}  {Naticchioni,}$^{6}$
{Antonio}  {Pasqualetti,}$^{10}$
 {Giovanni Piero}  {Pepe,}$^{1,2}$
{Maurizio}  {Perciballi,}$^{6}$
{Luca}  {Pesenti,}$^{8,9}$
{Paola}  {Puppo,}$^{6}$
{Piero}  {Rapagnani,}$^{5,6}$
{Fulvio}  {Ricci,}$^{5,6}$
{Luigi}  {Rosa,}$^{1,2}$
{Carlo}  {Rovelli,}$^{11,12,13}$
{Davide}  {Rozza,}$^{7,8}$
{Paolo}  {Ruggi,}$^{9}$
{Naurang}  {Saini,}$^{5,6}$
{Valeria}  {Sequino,}$^{1,2}$
{Valeria}  {Sipala,}$^{7,8}$
 {Daniela}  {Stornaiuolo,}$^{1,2}$
 {Francesco}  {Tafuri,}$^{1,2}$
{Arturo}  {Tagliacozzo,}$^{1,2,14}$\footnote{Corresponding author}
{Iara}  {Tosta e Melo,}$^{7,8}$
 {Lucia}  {Trozzo,}$^{2}$
 \vspace{1.0truecm}}

\affiliation
{ $^1$
Dipartimento di Fisica Ettore Pancini,  Universit\ax degli Studi di Napoli Federico II, 
Via Cinthia,  80126-Napoli,   Italy\\
$^2${Istituto Nazionale di Fisica Nucleare, Sezione di  Napoli},  { {Via Cinthia},  {80126-Napoli},    {Italy}}\\
$^3$ {Consiglio Nazionale Ricerche-Istituto Nazionale di Ottica (CNR-INO), }  { {Via Campi Flegrei, 34},  {80078-Pozzuoli},    {Italy}}\\
 $^4${Consiglio Nazionale Ricerche-Istituto ISASI},   { {Via Pietro Castellino, 111},  {80131-Napoli},   {Italy}}\\
 $^5${Dipartimento di Fisica},  {Universit\ax di Roma La Sapienza},  { {Piazzale Aldo Moro, 2},  {00185-Roma},     {Italy}}\\
 $^6${Istituto Nazionale Fisica Nucleare, Sez. Roma1},   { {Piazzale Aldo Moro, 2},  {00185-Roma},   {Italy}}\\
 $^7${Dipartimento di Fisica Giuseppe Occhialini, Universit\ax degli studi di Milano Bicocca},  { {Piazza della Scienza  3},  {20126-Milano}, {Italy}}\\
 $^8${Laboratori Nazionali del Sud},  {Istituto Nazionale Fisica Nucleare},  { {Via Santa Sofia, 62},  {95123-Catania},   {Italy}}\\
$^9${Dipartimento di Scienze Chimiche, Fisiche, Matematiche e Naturali},
 {Universit\ax di Sassari},  { {Via Vienna, 2},  {07100-Sassari}, {Italy}}\\
 $^{10}${European Gravitational Observatory - EGO},   { {Via Edoardo Amaldi},  {56021-Cascina},   {Italy}}\\
 $^{11}${Campus of Luminy},  {Centre de Physique Theorique},  { {Case 907},  {Marseille},  {F-13288},  {France}}\\
 $^{12}${Aix Marseille Universit\'e},  {Aix Marseille Universit\'e, CNRS, CPT, UMR 7332},   {Avenue Robert Schuman},  {Marseille},  {F-13288},   {France}\\
$^{13}${Universit\'e de Toulon},  {Universit\'e de Toulon, CNRS, CPT, UMR 7332 },  { {Avenue de L'Universit\'e},  {La Garde},  {F-83130},   {France}}\\
$^{14}$ CNR-SPIN, Monte S.Angelo via Cintia, 80126-Napoli, Italy
 \vspace{1.0truecm}}
 
\date{\today}

\begin{abstract}
We provide a novel methodological approach  to the  estimate  of the change of the Quantum Vacuum electromagnetic energy density   in a  High  critical Temperature  superconducting  metal bulk sample,  when it undergoes the  transition in temperature,  from the superconducting to the normal phase. The various contributions to the Casimir energy in the two phases are highlighted and compared. While the TM polarization of the vacuum mode allows for a macroscopic description of the superconducting transition,  the changes  in the TE vacuum mode induced by the superconductive correlations are introduced within a  microscopic model, which  does not explicitly take into account the anisotropic structure  of the material.
%
%
%
\end{abstract}

\maketitle

\section{Introduction}

 
  The electromagnetic (e.m.) field does work on each unit volume of matter at the rate $ \vec{E}\cdot \vec{j}$, where $\vec{E}$ is the electric field and $\vec{j}$ is the charge current density.  Feynman and coauthors, in their textbook on electromagnetism\cite{feynman} stress  the indefiniteness in the location of the e.m. field energy:``{\sl It is  sometimes claimed that this problem can be resolved by using the theory of gravitation... all energy is the source of gravitational attraction}". The Archimede project  is designed for measuring the effects of the gravitational field on a Casimir cavity by performing a weighing measurement of the vacuum fluctuation force on a rigid Casimir cavity\cite{avino,calloni,Allocca:2012kw}. 
The vacuum state of the  e.m. photon field is strongly modified in presence of a metal material, forming a coherent radiation-matter realm\cite{leger}.  The goal of this project is to measure changes in the Casimir force when the cavity  metal undergoes a phase transition from the normal metal  phase  to the superconducting  state. In the following we will address the two phases, by talking shortly of a normal metal or  a superconducting metal. There are speculations that the Casimir force can be the driving microscopic mechanism for superconducting pairing\cite{kempf}.  In this paper we adopt a more conservative view  and assume that the largest contribution to the  change in the Casimir force  at the transition comes from modifications  of  the vacuum fluctuation spectrum due to changes  in the  photon field density of states at long wavelength,  assuming that the thermodynamic free energy gain at the transition (the so called "condensation energy"), originates instead at atomic scale, by including short distance lattice effects. The latter are considered as a small correction  to the vacuum fluctuation spectrum and can be measured at very low temperature  with a  transition in magnetic field. 

This work is devoted to the  comparison in the Casimir energy between the normal and the superconducting phase of a metal slab considered as the Casimir cavity in free space. By choosing an High Temperature Superconductor  (HTS) as YBCO we gain various advantages.The transition temperature is relatively high, what increases the feasibility of the experiment.  We choose $\hat z$ in the direction  of the $c$ axis  orthogonal to the HTc  superconductor planes, so that  the collection of  $Cu O$ planes are parallel to the planar surfaces of the material, thus  exploiting  the strong anisotropy of the superconducting correlations. The dominant contribution to  the Casimir energy for a  normal metal  slab comes from the plasma modes that can be excited  at the opposite surfaces. Retardation implies that they are both acoustic with a top frequency in the crossover between  microwaves  and infrared radiation ($\sim$ THz), an energy range which is already rather high for a conventional superconductor.   In spite of the fact that the electronic spectrum is not fully gapped in the superconducting phase, coherence of a HTS is  expected to be more robust  and preferable in this range of frequencies. Ignoring in this approach the nodes in the gap,  YBCO has a maximum superconducting gap $\Delta\sim ${\sl  tens of  meV} which is in the same frequency range. 

 Differences arise between the superconductor and the normal  phase, because the minimal coupling of the e.m. field to the superconducting order parameter  generates the Anderson-Higgs (AH) mechanism in the superconducting state\cite{anderson0,anderson2,negele}.  The two transverse massless modes of the Maxwell equations in vacuum are replaced by three  independent massive modes with  mass $m^2 = ( 2\: \pi / \lambda _L)^2$  which, macroscopically,  gives rise to the Meissner effect, i.e. the expulsion of the static magnetic fields from the superconducting bulk.  This fixes an energy threshold for photon propagation inside the superconductor, given by $ \hbar c m/n$, where $n$ is the refraction index (denoted as Meissner threshold in the following). 
While the superconducting correlation length $\xi$ can be of the size of the sample,  the Meissner penetration length  is relatively small in the $c-$axis direction $\lambda _L^{\perp ,YBCO} \sim 0.75 \: \mu m $, where $ \lambda _L$ is the London penetration length. The latter is of the order of the skin depth  in the normal metal, at least for a pure sample. However, the  TE  vacuum modes, characterized by $E_z$ penetrating in the sample,  perform in any case very differently between  the two phases  as for what concerns  the interaction with the surface plasma excitations.  Resonant tunneling  below the Meissner threshold, assisted by virtual quasiparticle (qp) electronic excitations is still  possible if the slab has thickness $ a \gtrsim  2 \lambda_L$, as will be explained in the following.

 For a macroscopic metal body of linear  millimiter size  $a$,  in coherence with the e.m. vacuum,  a macroscopic approach  is usually adopted, resorting to a semiclassical response theory in terms of a dielectric function $\epsilon (\omega )$  (intended at $k\approx 0$ for an isotropic system). A macroscopic description of the TM photon vacuum in the presence of a slab-like  cavity is allowed as  the TM modes  have $B_z =0$  which  can be macroscopically compatible with the metal both in the normal and superconducting phase. Indeed superconductors  require $B_z=0$ at the boundary with  the  plane surface due to Meissner effect.   Among the  non vanishing field components ($E_x$, $E_z$ and $B_y$),  the  $ \epsilon (\omega) E_z $ component should be matched at the boundary. In a slab geometry $\epsilon (k_\parallel, \omega )$ entails plasma surface modes ($k_\parallel \parallel a-b$ planes)  coupled between the two opposite surfaces, which  can be classified as symmetric plasma mode (SPM)  and antisymmetric plasma mode (ASPM) with respect to the inversion plane of the slab. Only the  Transverse Magnetic (TM)  photons couple to these modes.   
 It is well known that  in the normal metal the SPM and the ASPM give opposite, almost compensating,  contributions to the Casimir energy and the APM prevails  with its minus sign\cite{bordagMath}.   We will argue that the superconductor has collective modes corresponding to the SPM and APM, the Mooij-Sch\"on (MS) mode and the Carlson Goldman (CG) mode,  respectively. 
On the other hand, in  the superconducting phase the MS  mode is a true plasma oscillation mode, while the CG one is macroscopically charge neutral, balancing a qp electron component with a Cooper pair  component  which, in the case of nodes in the gap, do not require too much energy. The CG mode  being neutral,    does not couple macroscopically to the zero point extended  TM photons, thus implying the absence of compensation occurring in the normal phase, what  makes  a sizeable difference  when comparing the results  of the two phases  in first approximation. Moreover, in the superconducting phase, the two transverse massive modes  are both similar to TM modes and they  both couple to the MS excitation mode. 
 
  The TE mode (which is characterized by non zero $ B_x,B_z,E_y$ components) does not couple to the plasmonic  surface excitations  in the normal metal, at least in a macroscopic approach based on a  Drude-like frequency dependent  dielectric  constant $\epsilon (\omega )$.   In the superconducting case, the longitudinal  AH massive photon mode is similar to a TE mode   and we expect that it does not couple with the MS plasmon, either. On the contrary, it allows for longitudinal resonant states to arise, which split off the minimum of the AH band  in the confined  geometry,  provided  Cooper pairs can be, even just virtually, broken. These pair broken states   couple with the $B_z$ component of the incoming AH mode.  However, to describe this physics, the macroscopic picture cannot be adopted. In fact, the  $B_z $  component  of  the longitudinal simil -TE mode would violate the  macroscopic London  $B_z =0$ boundary condition  for the superconductor. 
   A realistic model  accounting for the full microscopic electronic structure of the real sample  is beyond any possible approach.  The idea is to replace the actual cavity  with an  effective  local interaction between   the non vanishing  $B_z$ component and pair breaking in the $a-b$ planes.  For YBCO the Meissner threshold is of the order of the gap $\Delta$, so that the resonant state can be located  in the coherent subgap energy window. 
We will  adopt a two channel scattering approach for the TE mode,  by  considering virtual photon emission or absorption processes, as the result of the   interaction of the incoming wave  with quasiparticles close to the nodes of the gap.  We will show that   resonant states arise in the case of normal incidence  of the TE mode onto the film surfaces.    
   
  In Section II we present the macroscopic approach for deriving the contribution to the Casimir energy  from interaction of the TM mode  with the surface plasma waves for various linear lengths of the sample which plays the role of the Casimir cavity.  Section II.A  discusses the case of a normal medium, while Section II.B is devoted to the superconducting medium. The ideal normal metal is characterized by a single parameter, the plasma frequency $\omega _p$.
   Hence, the length scale  is $c/\omega _p$ where $c$ is the propagation velocity in the material. According to London theory, $\omega _p$ is replaced   in the superconducting phase by the superconductive plasma frequency  $\omega _{ps} =(4\pi n_s e^2 /m)^{1/2}$, in which the density of Cooper pairs $n_s$ appears  replacing the electron density $n$. Here  $c/\omega _{ps}$ is the London penetration length $\lambda_L$  for an isotropic medium. This implies that close to the transition temperature  the normal and the superconductive length scale are quite different, while for temperatures not in the transition region the two scales can be considered as being roughly equal.   In Section III we will present the  effective model for the microscopic model approach  of a TE mode characterized by the $B_z$  field component propagating at normal incidence. Details are given in Appendices A and B.   Section IV is devoted to the Casimir energy of the superconducting phase. Further scattering features of the model, including  phase shift jumps  are critically analyzed. The total Casimir energy for the two phases and their difference is presented in Section V.   Section VI  includes a summary and the conclusions that can be extracted which could be useful in the interpretation of the experiment. 
   
   As a final warning, any time we discuss qualitative physics related to superconductors we assume zero temperature and  ignore the fact that HTS materials, YBCO in particular, are strongly anisotropic and that  there are nodes in the  {\sl d-wave} excitation gap. In this sense, the gap is $2\Delta$  with no qp's in this energy range  both in the text and in the pictures. Also the velocity of light is  denoted as $c$ with no care of the refractive index. These  simplifications  aim to highlight the differences of the superconducting phase  with respect to the normal phase. We are  aware, of course,  that quantitative analysis would require to include these peculiarities of the HTS carefully, and we mention and introduce them in the text and in the numerical estimates, when they cannot be overlooked. 
   
 The approach considered in the present explicitly uses the bulk behaviour of the superconductor, while in the case of the experiment one can consider both the use of bulk samples and thin films, and also the superposition of thin layers. In this sense it is expected to extend this work to the limiting case of thicknesses tending to zero, in the nanometer limit.
   
\section{Comparison between the  normal and the superconducting phase energy scales}
To compare the superconducting and the normal phase of our sample from the macroscopic point of view, we have to define the  dielectric properties of the two phases with the energy scales involved. 

In the case of the normal phase, when the inelastic scattering time  $\tau $ is long  enough (i.e. in the limit  $\omega \tau >>1$),  we can assume that the sample is close to be an ideal metal. With the TM polarization, $E_z$ penetrates inside the metal over a length $\tilde\delta$ named 'skin penetration depth'.  The Drude conductivity  for  the ideal normal metal  can be used: 
    \beq
    \sigma = \frac{n \:e^2}{m}  \: \frac{1}{1/\tau -i \omega } ,  \:\:\: \to   \sigma _2 (\omega ) =  \sigma_0  \:\frac{\omega \tau}{ 1+ \omega ^2 \tau ^2} 
    \nonumber
    \eneq 
(with $\sigma _0= n e^2 \tau /m$) allows to define a frequency scale:
 \bea
  \omega _0  = \frac{c}{\delta } =c \: \sqrt{  \frac{4\pi }{c^2} \sigma _2 (\omega _0) \omega _0}  = 4\pi \: \sigma _2 (\omega _0). 
  \label{oppi}
  \enea 
For the ideal normal metal Eq.(\ref{oppi}) recovers   $\omega _p  = \sqrt{ 4\pi \: n e^2/{m }}$, the normal metal plasma frequency. 

\begin{figure}[h]
\begin{center}
\includegraphics[scale=0.6]{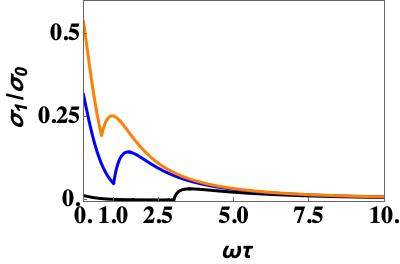}
\caption{ $\sigma _1$  vs $\omega  \tau $ for  $ \Delta\cdot  \tau = 1.5\: (lowest \: curve),0.5,0.3$ and  $ k_BT\cdot \tau = 0.3$. $\tau $ is the inelastic scattering time   } 
\label{sig}
\end{center}
\end{figure}

\begin{figure}[h]
\begin{center}
\includegraphics[scale=0.5]{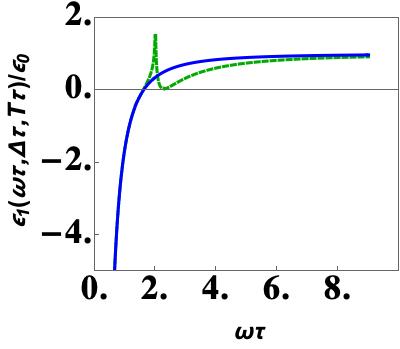}
\caption{comparison between   $\epsilon_1(\omega\tau, \Delta \tau, T  \tau  ) $  of the normal ideal metal case  $\epsilon_1(\omega\tau, 0, T  \tau  ) =1-2.6/( \omega \tau)^2$ ( {\sl full blue curve}) and  for the superconducting case, $\epsilon_1(\omega\tau, \Delta \tau, T  \tau  ) $   ( {\sl orange dashed curve})   vs $\omega  \tau $.  Here   $ \Delta\cdot  \tau = 1.$ and  $ T\cdot \tau = 2.0$,  (with $\sigma_0 = 0.7$). The sharp peak at $ \omega \sim 2\Delta /\hbar$ heralds the enhancement of qp excitations at the pair breaking energy. Difference between the two curves is only close to  $\omega \sim 2 \Delta/\hbar$.  } 
\label{ridifo}
\end{center}
\end{figure}

With the chosen geometry, the nodal lines of  the   d-wave order parameter for HTc superconductors lie  in the $a-b$ plane parallel to the surfaces of the material. Although the nodes of the gap  $\Delta $ imply that some density of qp's is excited  even at $T\approx 0$,  we consider the gap $\Delta $ quite robust for transport in the $c$ direction. 
 In Fig.\ref{sig} we report the real part of the conductivity, $ \sigma _1 $, at finite frequency, at $\vec{q} =0$ (i.e. its bulk value),  for increasing $\Delta $ at fixed temperature  for an $s-wave$ superconductor, as derived from Ref[\onlinecite{berlinsky}]. When $k_BT << \Delta$,  $ \sigma _1 $ is quite small at  frequencies  $\omega  < 2 \Delta$, except for  the pseudo Drude peak at zero frequency, which contributes to the sum rule $  \int _0^\infty d\omega \: 4\pi \sigma_1(\omega ) = \pi \: \omega_{ps}^2/2$ in the limit $\tau \to \infty$   and is not included in the plot. This implies that  the Kramers-Kronig  transform for  $ \sigma _2 $,
\beq
\sigma_2(\omega ) = - \frac{\omega}{\pi} \int_{-\infty}^{\infty} d\omega '  \: \frac{ \sigma_1 (\omega ')}{{\omega '}^2-\omega ^2}.
\label{opi}
\eneq
 is dominated mostly by the enhancement of excitations close to the pair breaking energy, but  also by  the zero   frequency peak. Including just the latter $\delta -$peak,  we obtain    
 \bea
  \sigma _1 <<  \sigma_2 \sim \frac{ n_se^2 }{m \omega },
  \label{pola}
  \enea
  where $n_s$ is the Cooper pair density.
   Hence, the polarizability is quite high at microwave frequencies\cite{glover}  
 \beq
 \epsilon_1(\omega ) \approx - 4\pi \sigma _2 /\omega,
 \label{glov}
 \eneq
 in the limit $\tau \to \infty$. It follows that the superconducting dielectric function is not  much different from the normal one, except for frequencies close to the pair breaking energy $2\Delta/\hbar$, where it  has an enhancement just above  the gap threshold\cite{bardeenmattis}. 
  The main differences are expected in the quantitative energy scale of the modes and in their  lifetime\cite{holczer,klein,jin,steinberg}. 
   A plot of an approximation of the  real part $\epsilon_1(\omega ) $ for the superconducting and normal phase is displayed in Fig.\ref{ridifo}.


In the superconducting case, inserting  Eq.(\ref{pola}) in   Eq.(\ref{oppi}) we obtain  the superconducting plasma frequency,  $\omega _{ps}  = \sqrt{ 4\pi \: n_s e^2/{m }}$. 
 As expected,  the definition of Eq.(\ref{oppi}) appears  convincing on the full range $normal \:\leftrightarrow\:\: superconductor$, when screening is low. In the case of YBCO, the anisotropy of the London length is important as  $\lambda _L^\perp /\lambda _L^\parallel =5$ (where $\lambda _L^\parallel $ is in the $a-b$ plane).   With the choice  $\omega _{ps}   = \frac{c}{\sqrt{\epsilon_s} \lambda _L^\perp}$, which is valid in the limit $\omega \tau >>1$, and with     $ \epsilon _s = 30$, $\lambda _L^\perp  \sim 0.75 \: \mu m $, which gives  $ \omega _{ps} \sim 0.7 \cdot 10^{14} \: Hz = 48 m \: eV$.  The maximum gap for YBCO is $\Delta  \sim 16 m\: eV$, so that $2\Delta < \omega _p$, but rather close to it. 
  \section{ Photon modes in interaction with the metal film}
  \subsection{ TM polarization: Contribution of  the  plasma modes to the Casimir energy,  in the normal metal phase}
     Let us now introduce the Maxwell equation outside of the superconductor and the boundary conditions:
           \bea
 \left ( \nabla ^2 + \mu \epsilon \frac{\omega^2 }{c^2} \right ) \left \{ \begin{array}{c} \vec E\\\vec B \end{array} \right \} = 0. 
 \label{max}
 \enea
 The modes have dispersion $ \frac{\omega ^2}{c^2} = k_\parallel ^2 + k_z^2$. 
As the TM mode  has $B_z =0$,  it can be macroscopically compatible with the metal both in the normal and in the superconducting phase. Indeed superconductors  require $B_z=0$ at the boundary with  the  plane surface. Among the  non vanishing field components ( $E_x$, $E_z$ and $B_y$), continuity of $  D_z = \epsilon(\omega ) \: E_z $  is required, assuming  vanishing charge density at the surface.  These conditions, written for the TM mode component $E_z$  across a single vacuum-material boundary are:
 \bea
\epsilon_L(\omega ) \: \Phi _L -  \epsilon_R(\omega ) \: \Phi_R =0,\:\:\:\:\: &\Phi '_L - \Phi'_R =0,\:\:\:\: (TM).
\label{TMmode}
\enea
where $L$ is the vacuum  at the left  hand side, with  $\epsilon_L= \epsilon _0=1$, and $R$ is the material at the right  hand side with dielectric function $ \epsilon_R(\omega )$.  At microwave frequencies, the  electric field penetrates into the bulk  of the normal material over the skin depth\cite{skin}, which is 
of the order of the London penetration length  at these frequencies. Bound states at surfaces $z=0$ and $z= a$ imply: 
\bea
 \Phi = \left \{ \begin{array}{cc}  A \: e^{\kappa_0 z } & z < 0 \\  B  \: e^{-\kappa_a z } + C  \: e^{-\kappa_a  (a-z) } &0<  z < a \\  D \: e^{-\kappa_0 z } & a< z  \end{array} \right . ,
 \label{tre}
 \enea 
with 
\bea
   \freccia      \epsilon_{a/0} (\omega )  \: \frac{\omega ^2}{c^2}= k_\parallel ^2 - \kappa _{a/0} ^2,  
 \enea 
 where  the subscript $a$ refers to the $0<z<a$ space region, while  the subscript $0$ is for the vacuum regions. 
Assuming inversion symmetry at  $a/2$ is   $ C = \pm B $ and  $D = \mp A $.  Hence, Eq.s(\ref{TMmode}) at  $z=0$ become
\bea
\left \{ \begin{array}{c} B \left [ - \kappa_a  \pm \kappa_a e^{-\kappa_a a } \right ] - A \: \kappa _0 =0 \nonumber\\
\epsilon (\omega )  \: B \left [1 \pm e^{-\kappa_a a } \right ] - A  =0 \nonumber  \end{array} \right .  
\enea
The requirement $ det_\mp =0 $ implies: 
\bea
  -\frac{ \kappa_a }{\kappa_0} \left [ 1 \mp e^{-\kappa_a a} \right ] =\epsilon_a (\omega )  \:  \left [1 \pm e^{-\kappa_a a } \right ] 
\label{quat}
\enea
  In the case of no retardation and ideal metal, the two plasma modes are very simple. No retardation ($ c\to \infty $) implies   $ \kappa _a = \kappa_0 = k_\parallel$. Adopting the Drude form for  $\epsilon _a (\omega)= 1- \omega _p^2/\omega ^2 $ where  $ \omega _p^2 = 4\pi n \: e^2/m $  is the plasma frequency for electronic density $n$, electron charge $e$ and  mass  $m$, we get:  
   \bea
 det _+=0  \freccia   \omega_+ ^2 = \frac{ \omega _p^2}{ 1+ \coth{\left (\frac{ k_\parallel a}{2}\right ) }}\nonumber\\
 \enea 
 for the symmetric  plasmon and 
 \bea
  det _-  =0  \freccia   \omega_- ^2 = \frac{ \omega _p^2}{ 1+ \tanh{\left (\frac{ k_\parallel a}{2}\right ) }}, 
  \label{quat2}
 \enea
 for the antisymmetric plasmon.   $  \omega_+( k_\parallel a)$ is an acoustic mode and is rather unchanged when retardation is included, while  $  \omega_-( k_\parallel a )$  is strongly modified by retardation. 
   
 Defining $ f_+ [x] = \coth[x] $ ($ f_- [x] = \tanh[x] $)  the total energy associated to each plasma mode is: 
We now integrate on $k_\parallel$.
\bea
{\cal{E}}_\pm^{TM}
 \equiv\met \sum _{k_\parallel}  \omega _{k_\parallel}^\pm \nonumber\\
  = \met \frac{L^2}{(2\pi)^2} \int_0^{+\infty}  2\pi \: k_{\parallel} \: dk_{\parallel} \frac{ \omega _p}{ \sqrt{ 1+ f_\pm \left [\frac{ k_\parallel a }{2}\right ] } }
\enea
or, with   $ k_{\parallel}L = x $, 
:

\bea
 {\cal{E}}_\pm^{TM}
 \equiv \frac{ \omega _p}{2\pi} \int_0^{+\infty} x\: d x \frac{1}{ \sqrt{ 1+ f_\pm \left [ \frac{ a  x}{2L}\right ] } } \nonumber\\
=    \frac{ \omega _p}{2\pi}  \:  \left (  \frac{2L}{a } \right )^2 \: \met \left \{ - \int_0^{+\infty}  d y\:\frac{y^2}{2} \: \frac{\partial }{\partial y} \frac{1}{ \sqrt{ 1+ f_\pm \left [y\right ] } } \right \}.
\label{noret}
\enea



To include  retardation,
 we define $  \kappa =  \sqrt{   k_{\parallel} ^2- \frac{ \omega ^2 } {c^2  }  \epsilon(\omega)}$ and  use $\lambda  = a  \frac{\omega _p}{c}$, $ s =k_\parallel a $, $ \nu  = \frac{\omega }{\omega _p}$.   
  \bea
 0= det _\pm  = \sqrt{ \frac{ s^2+\lambda^2 (1- \nu^2)}{ s^2-\lambda^2 \nu^2 } } \nonumber\\
 +\left ( 1-  \frac{1}{ \nu^2}\right ) \: f_\pm \left [{\frac{\sqrt{ s^2+\lambda^2 (1- \nu^2)}}{2}}\right ]. 
     \label{impl}
   \enea
   Solutions are only for  $s > \lambda \nu \to k_\parallel > \omega /c$. This limitation guarantees that they are bound states (decaying outside the slab).  
 At very small $s$, $\nu$  has to be also very small to keep the square root  real. This implies that $ \left ( 1-\frac{1}{\nu^2} \right ) $ is strongly negative  and a  a solution is always found  for both 
 equations.
 \begin{figure}
 \centering
\def\big{\includegraphics[height=6.8cm]{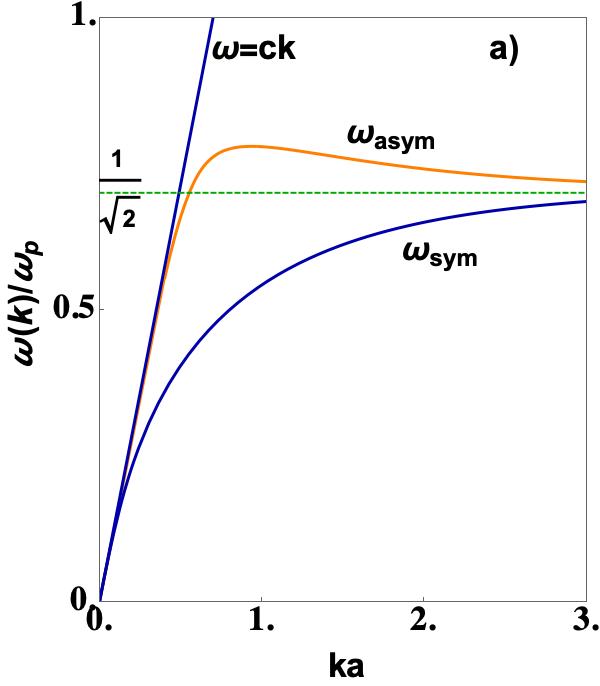}}
\def\little{\includegraphics[height=3.2cm]{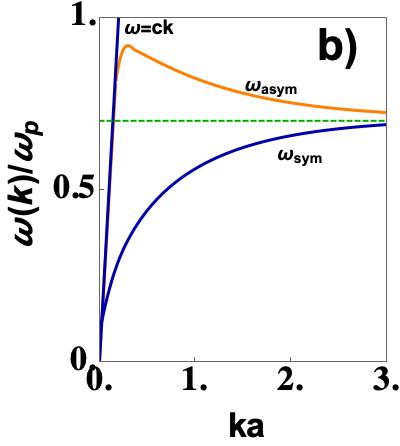}}
\def\stackalignment{l}
\topinset{\little}{\big}{+78pt}{+78pt}
\caption{ The symmetric ($ \omega _{sym}$) and antisymmetric ($ \omega_{asym}$) dispersion  relation for the plasma bound excitations   vs. $ k_\parallel$ (with $ k_\parallel > \frac{\omega}{c}$), in the retarded normal case, for two different  linear sizes of the sample, $a\omega _p/c =0.7$ $(a)$ and  $ a\omega _p/c =0.2$ {\sl inset  ($b)$}). The linear  light dispersion $ \omega = ck$ has also  reported in the corresponding scale.   The {\sl dashed green  line} is the asymptotic limit at $ \omega _p/ \sqrt{2}$. } 
\label{dispersion}
\end{figure}
We define $ \nu_\pm  ( s) $ is the solution of $det_\pm  ( s,\nu, \lambda ) =0 $  and is given in Fig.\ref{dispersion} for $\lambda = 0.7$  $(a)$ and $0.2$  ($(b)$,{\sl inset}).  The function $det_\pm  [s,\nu,\lambda] $ depends on  $a  $. Dropping the label $\pm$ for simplicity we get 
  \begin{figure}
      \includegraphics[height=90mm]{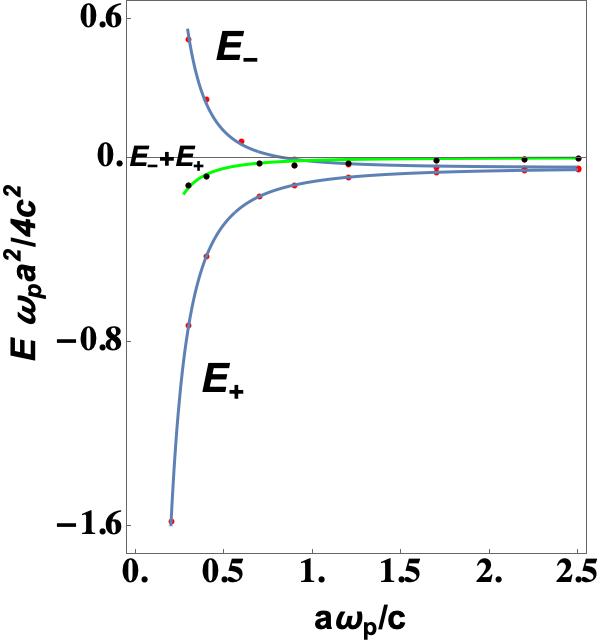}
\caption{   contribution to the  Casimir energy, per unit area $L^2$,  for various linear sizes $a c/\omega_p$, coming from the plasma modes in the normal ideal metal sample.  $E_+ (E_-)$ comes from the symmetric  (antisymmetric) plasma mode. The {\sl full  blue curves} are fits  of the {\sl red dots}  with  $ z^{-\alpha}$,  where $\alpha \approx 5/2 $ for $E_-$ and  $\alpha \approx 2 $  for  $E_+$. $E_-$ is negligible  beyond   $d c/\omega_p \sim 1 $. The {\sl black dots} correspond to  the sum   $E_++E_-$.  The fit  {\sl green curve} is a power law  just weaker than  $ z^{-3} $ ( see Section III.D).  } 
   \label{vanda}
\end{figure}

\bea 
{\cal{E}}^{TM} \equiv \met \sum _{k_\parallel}  \omega _{k_\parallel}=   \frac{ \omega _p}{2\pi}  \:  \left ( \frac{2 L}{a} \right )^2  \nonumber\\ \times \met  \left \{ - \int_0^{+\infty}  d s \:\frac{s^2}{2} \: \int d\nu \: \frac{\partial  \nu }{\partial s}   \: \delta \left ( \nu -\nu ( s) \right ) \right \} .
\label{no1}
\enea
Here $  \frac{\partial  \nu }{\partial s} = - \left . \frac{\partial\:  det }{\partial s} /{\frac{\partial \: det  }{\partial \nu}} \right |_\lambda  $.

For the case $a \omega _p/c = 0.2 $ we have: 
   \bea
     {\cal{E}}_\alpha^{TM}  \equiv  \eta _{\alpha} \times     \frac{ \omega _p}{2 \pi}  \:  \left (  \frac{2 L}{a} \right )^2 ,\label{epla} \\
   {{\eta}}_{-}^{ret}  \approx  0.157931,\:\nonumber\\
  {{\eta}}_{+}^{ret} \approx  -0.204317,  \nonumber
\enea
(label $\alpha = \mp$ stand for antisymmetric and symmetric, respectively) to be compared with the non retarded ones 
from Eq.(\ref{noret}),   ${{\eta}}_{-}^{noret}  \approx  0.167578,\: 
  {{\eta}}_{+}^{noret} \approx  -0.192115 $. While the energy dispersion $ \omega _- (k_\parallel)$  is quite different for small $k_\parallel$ and becomes acoustic for both modes in the retarded case,  
the difference in their contribution to the  Casimir energy, per unit cross section  area $L^2$, is rather  small. This is shown in Fig.\ref{vanda}   for various linear  widths of the sample $z= a c/\omega_p$. $E_\pm$ and their sum   $E_++E_-$ ({\sl black dots}) are reported vs. $a c/\omega_p$. The  fitting of the  sum  $E_++E_-$({\sl green curve}) gives   a scaling  law close to $ z^{-3} $  but weaker  (see discussion in Section III.C).

\subsection{TM polarization: plasma excitations in the superconducting phase}

 Because of the presence of the gap $\Delta $ in the spectrum,  it can be argued that the plasma  modes are much better defined in the superconducting case  than in the normal phase because, with the exclusion of the nodes of the energy excitation spectrum,  in the rest of the $2-d$ Brillouin zone, they are located in the energy gap.  In the case of the TM mode,   
continuity of  $ \epsilon  (\omega ) E_z $ at the boundary with exponential decay  inside the sample provides  the energy dispersion  of the plasma modes. 

  The  plasma mode dispersions  for a superconducting film have been  plotted with a "phenomenological" approximation in Ref.[\onlinecite{doria1}]  and they do not look much   different from our Fig.\ref{dispersion}, except for the frequency scale which replaces the normal metal plasma frequency $\omega _{p}$ with the superconducting one,  $\omega _{ps} $ (the $2-d$ mode dispersion  implies in both cases an extra factor $1/\sqrt{2}$). 
   Similarly to the normal case of Fig.\ref{dispersion} the symmetric and antisymmetric modes  of Ref[\onlinecite{doria1}] are $\sim \sqrt{k_\parallel}$ and   $ \sim 1/  \sqrt{k_\parallel}$  respectively.
 However,  the similarity is conceptually misleading.  In the case  of  Ref[\onlinecite{doria1}], the film is embedded in a non conducting medium with  an enormous value for the static dielectric constant ($ \tilde{\epsilon }\sim 2\cdot 10^4$). The phase velocity   of  Ref[\onlinecite{doria1}] is  $\omega / k_\parallel < c/ \sqrt{\tilde{\epsilon }}$, so that a  frequency independent approach  can be adopted.  This is equivalent to ignore retardation which is mostly relevant in our case, as  the film is located in vacuum and the bulk plasma frequency  $\omega _{ps} \gtrsim 2\Delta $ , or   even  $ \lesssim 2\Delta $. 
More generally, we can estimate  the scale of the dispersion,  $\omega _0$,  for YBCO  as  $\omega _0 = R_{sq} / L $   where  $ R_{sq} $ is   the sheet resistance in the non-superconducting state  ( $\sim 2 \times 10^{-4} \Omega$ at $ \sim 10 GHz$) and $L$ is  the kinetic inductance (expected to be much larger than the geometric inductance)   (with  dimensions $\Omega \: sec $ and $\Omega $ is Ohm). In the case of an  $s-wave$ superconductor, the commonly used expression\cite{annunziata} is : 
 \bea 
L\approx \frac{R_{sq} \hbar}{\pi \Delta } \frac{1}{ \tanh \left (\frac{\Delta}{2 k_BT_c} \right )}.  
\enea
Adopting this definition for $L$, and a BCS form for $\xi \sim 10 \: nm$,  we get
\bea
\omega _0 \sim  \frac{ v_F}{\xi} \sim 10^{14} sec^{-1}
\label{omeg0}
\enea 
In the limit  $\omega \tau >>1$ the [inductance  per unit length$]^{-1} $ for a  sample  of $1 \: \mu m $ length can be  estimated as $ n_s e^2 /m \sim 10^{19}  (  \Omega \times sec )^{-1} $, which produces the same order of magnitude for $\omega _0$  as Eq.(\ref{omeg0}).

 \begin{figure}[h]
\begin{center}
\includegraphics[scale=0.47]{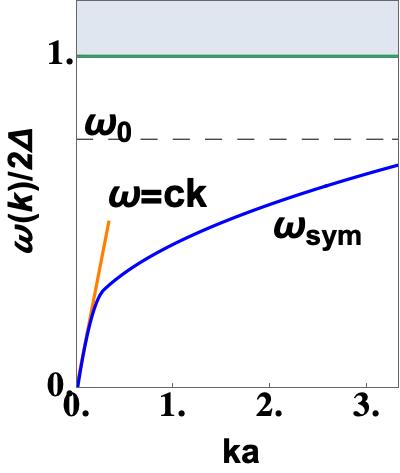}
\caption{ The dispersion  of the symmetric Mooij-Sch\"on mode   ($\propto   \sqrt{k} $)  in a superconducting film. Vertical axis not in scale} 
\label{mooij}
\end{center}
\end{figure}

 Mooij and Sch\"on\cite{mooij} (MS)  derived the collective excitation modes in reduced geometries  from an hydrodynamical approach for charge imbalance.  In  a superconducting wire  of diameter  $r_0$ a  linear dispersion  mode is well defined for $ k r_0 << 1$, with  velocity  $ c_{pp} ^2  \approx   \omega _{ps}^2 \frac{r_o^2}{2} \epsilon_s ^{-1} \ln ( 1/ k r_o)  $, where $\epsilon_s $ is the dielectric constant in the superconducting phase. 
      They also remark that, in the case of a superconducting  slab of thickness $d$  in vacuum, the screening voltage  is $ \delta V( x, \underline{r})$ with  $\underline{r}$  in the surface plane. The 2d Fourier transform is $\delta V =\frac{2\pi}{k} \: \tilde\sigma   $, where  $ \tilde\sigma =\delta  \rho_s\: d $ is  the induced surface charge density  ($\delta  \rho_s $ is the volume  induced  charge and $d$ is the thickness of  the slab in the third direction).  The $3-d$  Fourier transform is $\delta V =  \frac{4\pi}{k^2} \rho $,  so that, we have 
 \bea
 \delta V ( k, a) =  \frac{2\pi}{k}  \left [ \frac{ a}{\epsilon_s} + \frac{2}{k} \right ] \: \delta \rho_s. 
 \label{scr}
 \enea 
 The continuity equation for the superconducting induced charge  with  $ \dot \rho_s = -i\omega F \rho_s  $ and   $F= 1+ 2\epsilon _s /( ka) $ given by  Eq.(\ref{scr}), together with Euler equation provides 
 \bea 
 \left ( \omega ^2 +i\: \omega \tau ^{-1}_{imp} \right ) F(k_\parallel) ={\omega '}_0^2
 \enea
 Keeping just the real part,  in the limit $ka <<1$ we have 
 \beq
 {\omega'} ^2 \frac{ 2\epsilon_s}{k_\parallel a} =  {\omega '}_0^2  \freccia \omega ^2 = \frac{ {\omega '} _0^2 a}{2\epsilon_s}\:  k_\parallel  
 \label{disdis}
 \eneq
 which is the MS acoustic mode for a slab,   with  $\omega \propto \sqrt{k_\parallel}$. It follows he $ k_\parallel$ dependence is the same  as of the symmetric plasma mode of the normal phase  of Fig.\ref{dispersion} and  of the  symmetric one  in Ref.[\onlinecite{doria1}], though not in scale.  It is reasonable to assume that $\omega _0'\sim \omega _0$ given in Eq.(\ref{omeg0}), so that the prefactor in the dispersion of  Eq.(\ref{disdis}) is of the order of:


 \beq
\sqrt{\frac{ \omega _{ps}^2 a}{2\epsilon_s}}  \sim 0.07 \div 0.3  \times 10^{14} m^{1/2} /sec ,
\label{disval}
\eneq
for  $ \omega _{ps} \sim 0.7\times 10^{14} Hz = 46 \: m eV $. The upper threshold for the  MS acoustic  mode   is  $\omega_0 < 2 \Delta \lesssim  \hbar \omega _{ps}$.

The mode corresponding to the ASPM is most probably the Carlson Goldman (CG) mode, which is close to the pair breaking  energy  and involves charge compensation between the  charge modulation  of the pair condensate  and the  charge modulation of the qp's.  At low temperatures, the CG velocity is $  c_{CG}$:
\beq
  c^2_{CG}  = \frac{n_s}{m} \frac{ 1}{2 N(0) } \approx  \frac{n_s}{n} \frac{ 1}{6}v_F^2
  \label{ccg}
  \eneq
 ($N(0)$ is the density of states at the Fermi energy). It  is expected to be quite short lived, particularly at small $k_\parallel$. Besides, being  this mode   charge neutral, it does not couple,  at first order, with the photon of the e.m. vacuum.
 The  signature of the pair breaking processes in the dielectric function appears at  about $ k_0 = \frac{2\Delta }{\hbar v } \sim  250 \: (\mu m)^{-1}$ , as discussed in Appendix C.   This   $k-$vector  refers to sampling distances of the order of the lattice spacing, beyond the validity of our approach.

      
 
  To sum up  the case of the TM modes, our  conclusion  is that,  in first approximation,  the CG neutral mode  which corresponds to the ASPM  in the normal metal case does not contribute to the Casimir energy because it does not couple to the zero point photon field. The  SPM  instead,  grows linearly with  $ k_\parallel $ at low  $k-$vectors  and bends as $ \sqrt{ k_\parallel }$,   not different from the normal ideal metal, but with an energy scale which  is different from the normal case,  given by  Eq.(\ref{omeg0}). The SPM contributes to lower the Casimir energy. In the superconducting case here is no subtraction of the positive contribution given by the ASPM, as it happens for  the normal metal TM case.  In the next Section we discuss the TE case for the superconducting metal, in which   resonant propagating modes may be present  below the AH threshold.
  
  \subsection{total Casimir energy in the normal phase}

\begin{figure}[h]
\begin{center}
\includegraphics[scale=0.4]{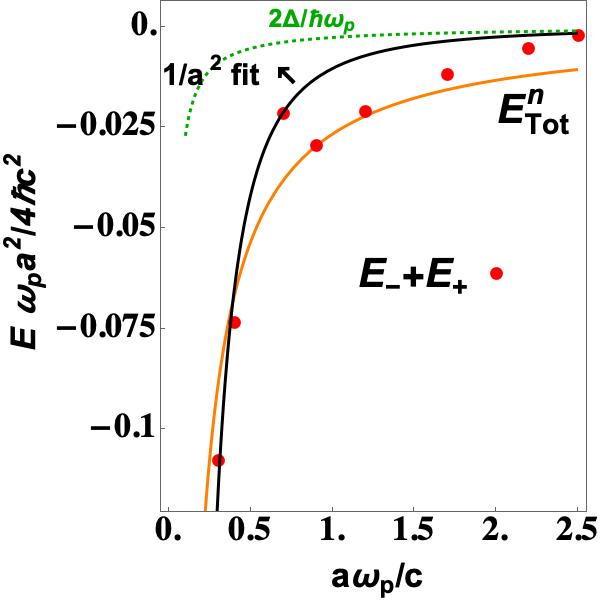}
\caption{  Total energies  per unit surface,   $E_++E_-$ of Fig.\ref{vanda} ({\sl  red dots})  and  $E_{Tot}^n$ evaluated at $\kappa d = 0.05$   vs. the  linear size of the sample $ z= a\omega_p/c$. $E_{Tot}^n$  scales with the linear size of the sample as  $z^{-3}$ and has been shifted  to match  with $E_++E_-$ at $ z=1.$. The weight of the change in the density of states due to the propagating states increases at larger $a$ values. A fit $ E\: a^2 \sim 1/a^2$ ({\sl green curve}), has been added, as well as the pair breaking threshold $2\Delta / \omega _p$ in the superconducting phase,  for comparison.      } 

\label{totSumo}
\end{center}
\end{figure}

\begin{figure}[h]
\begin{center}
\includegraphics[scale=0.4]{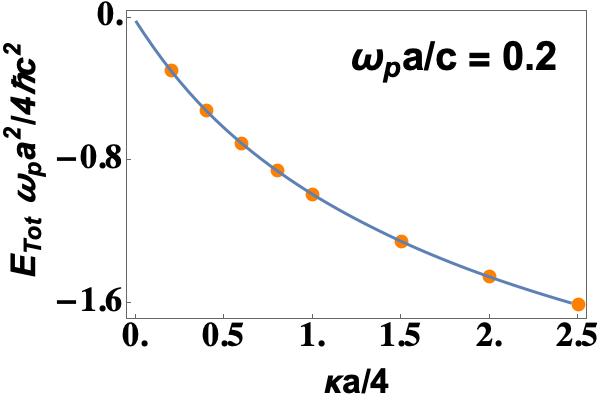}
\caption{  Total energy  $E^n_{Tot}$ per unit surface from  Eq.(\ref{totto}), at various potential strengths $\kappa a$ for  the  linear  width of the sample $ a c/\omega_p =0.2$. A   constant prefactor has been adjusted to scale the amplitude of the result  at values  corresponding  to Fig.\ref{totSumo}, when  $ \kappa d  =0.005$.    } 
\label{tot}
\end{center}
\end{figure}

 An estimate of  the Casimir energy for the normal phase of the sample requires the full  density of states of the photon propagating modes at energies which correspond to the Meissner window of the superconducting phase.
 These energies contribute to the   total Casimir energy difference, from the normal phase side.  A tutorial approach to this contribution  can be envisaged by adopting a  simple model for the transmission across the sample.   In this case a single elastic channel suffices because qp's  in the metal only contribute to the propagation  with a finite lifetime. Following Bordag\cite{bordagMath}, we mimick  the cavity as in Section IV, with two $\delta -$ function potentials at  the distance $2d$. The zero point energy of a photon  of wavevector $ K = \sqrt{\left |k_\parallel^2+k^2\right |}$ and energy $ \met  \hbar c K$, where $c$ is the velocity  of the incoming and outgoing photon in the vacuum. Scattering is assumed to be elastic. The strength of the $\delta-$functions is tuned by the inverse decaying length $\kappa$  of the field.  We anticipate here some results of a two channel scattering model that is presented in Section IV.B. The total transmission is 

\bea
t\left (k_\parallel,  k\right ) = \frac{ e^{i (k-q) d}}{1+ \frac{t_R}{t_R^*} r_R^* r_L \: e^{2 i q d}}  \nonumber\\
\to   \frac{1}{\left (1- \frac{k_\parallel^2}{k^2}+ \frac{ i\kappa }{2k} \right)^2} \frac{ e^{i (k-q) d}}{1- \frac{ \kappa^2}{4k^2} \frac{ 1}{\left (1- \frac{k_\parallel^2}{k^2}+ \frac{ i\kappa }{2k} \right)^2} \: e^{2 i q d}},
\label{trasm}
\enea  
where $k$ is the $k-$vector in the $\hat z$ direction out of the scattering region and  $q$ is the corresponding $k-$vector  between the two barriers (we take $q=k$).  $t_i \: (r_i)$ $( i=R,L$)  are the transmission (reflection) coefficients of the two $\delta -$potentials  which we have chosen equal.  This can also be derived restricting    the matrix $S_{12}$  of Eq.(\ref{matamat})  to a single channel. The total energy contribution coming from these delocalized states is\cite{bordagRep}:
\bea 
E_{tot}^n =  \met \sum _{k_\parallel}  \omega^{TM} _{k_\parallel}+  \:  \frac{L^2}{2} \!\!\int \!\!\frac{ 2 \pi k_\parallel dk_\parallel}{(2\pi)^2}  \nonumber\\
\times  \:c  \sum _\alpha   \sqrt{\left |\frac{\omega_\alpha^2}{c^2}+k_\parallel^2 -
\epsilon(\omega_\alpha) \frac{\omega_\alpha ^2}{c^2} \right |}\: \left . \left [ \frac{ \partial}{\partial k} \ln \frac{t(k_\parallel,k)}{t(k_\parallel,-k)}\right]\right |_{\omega _\alpha}.
\label{totto}
\enea
$ \omega^{TM} _{k_\parallel}$ are the plasma energy modes  in the first term arise from the poles of $t\left (k_\parallel,  k\right ) $.  $\omega _\alpha $ are the eigenvalues of the operator$ - \frac{d^2}{dz^2} + V(z)$  arising from the Schr\"odinger equation of the potential in the $z-$direction. The ratio    $\frac{t(k_\parallel,k)}{t(k_\parallel,-k)} = e^{2 i \delta }$ where $ \delta $ is the phase shift in the transmission. To subtract non-distance dependent terms from the expression of Eq.(\ref{totto}), we substitute $ t \to t /t_{d=\infty}$.  As the approach is only qualitative, we  rewrite it in the continuum limit  $\omega _\alpha  = c\: k$. 
 We get:
\bea 
E_{tot}^n =  \met \sum _{k_\parallel}  \omega^{TM} _{k_\parallel}+  \nonumber\\ \:  \frac{L^2}{2} \!\!\int \!\!\frac{ 2 \pi k_\parallel dk_\parallel}{(2\pi)^2}  \:c  \!\int_{0}^{+\infty} \!\!\!\! \frac{d k}{2\pi i}  \sqrt{\left |k_\parallel^2+k^2\right |}\:  \frac{ \partial}{\partial k} \ln \frac{t(k_\parallel,k)}{t(k_\parallel,-k)}.
\label{totto}
\enea By Cauchy theorem the $k-$integration  can be performed along the imaginary axis $ k \to i k$ and the deformation of the circuit shows that this integral already includes   the residues at the plasma  poles, so that the integral along the imaginary axis  provides the full contribution to the Casimir energy.  Integrating by parts, we obtain: 
\bea 
E_{tot}^n = - \frac{L^2}{2} \: \int \frac{ 2\pi \: k_\parallel dk_\parallel}{(2\pi)^2}  \hspace{2cm}\label{totto1}\\
\times  \: c \int_{0}^{+\infty} \!\! \frac{d k}{\pi } \frac{k}{ \sqrt{\left |k_\parallel^2-k^2\right |}}\: \ln  \frac{1}{\left |1-\left ( \frac{\frac{\kappa k}{2 }}{k_\parallel^2 -k ^2 - \frac{\kappa k}{2} }\right )^2 \: e^{-2  k d}\right |}. \nonumber
\enea
When $k_\parallel \gtrsim \kappa $, $t\left (k_\parallel,  k\right )$of Eq.(\ref{trasm})   has three poles  with increasing  $k$, which qualitatively reproduce the crossings with  the SPM curve, the  ASPM  and  the light dispersion  curve $ \omega = c k_\parallel$.  Their contribution to the integral is negative, positive and negative respectively, as expected, but there is no correspondence of the  location in energy with the dispersion laws of Fig.\ref{dispersion} . Rewriting  Eq.(\ref{totto1}) in dimensionless variables,  $ s' =2kd, s=2k_ \parallel d, \kappa'  = \kappa d$,  Eq.(\ref{totto1}) becomes, with $4d=a$:
\bea
E_{tot}^n =- \met \left (\frac{ 2L}{ a } \right )^2\:   \frac{\omega _p}{2\pi}  \times   \frac{2}{\pi} \frac{c}{  \omega _p a} \int s\:  ds \nonumber\\
\times  \int ds'  \frac{s'}{ \sqrt{\left |s^2-{s'}^2\right |}}\: \ln  \frac{1}{\left |1-\left ( \frac{\kappa ' s' }{s^2 -{s'} ^2 - \kappa '  s'}\right )^2 \: e^{-s'}\right |},
\enea
to be compared with the prefactor in Eq.(\ref{noret}). The  $\sim  z^{-3} $ dependence on the   linear  widths of the sample $z= a c/\omega_p$  is apparent.   At very small $\kappa$'s,  transmission is close to unity  for $k_\parallel=0$  and we expect  that $E_{tot}^n$ is roughly given by the plasma modes contribution only. It follows that  $E_{tot}^n $ should  be very close to the behaviour of $E_++E_-$ plotted  in Fig.\ref{vanda} ({\sl  red dots and green line}).  A numerical evaluation of the double integral at $ \kappa d  =0.005$ gives 
0.0075 and $E_{tot}^n $ does not match with $E_++E_-$  at  small $z = a\omega_p/c$.  However the two derivations  stem from different approaches and it is not of a surprise that the two results  do not match.   As the present approach cannot be considered  quantitatively faithful, we scale  $E_{tot}^n$ at $ \kappa d  =0.005$ to make it coincide with   $E_++E_-$  at  $z =1.0$.   In Fig.\ref{totSumo} ,   $E_++E_-$   vs $z$  is reported ({\sl  red dots}), together with  $E_{tot}^n  \approx  2\pi \: 0.0075/z^3$ ({\sl  blue curve}) and another fit $ \sim 1/z^4$ ({\sl green curve}).   At larger sample linear sizes  the weight of the propagating states increases and it is attractive, while the role of the plasma states  decreases, so that  derivative of   $E_{tot}^n$, the Casimir force, decreases. 
In Fig.\ref{tot} we plot  $E^n_{Tot}$, per unit surface,  derived from  the scattering model of Eq.(\ref{totto}) for  the  linear  width of the sample $ a c/\omega_p =0.2$, at various potential strengths $\kappa a$. A   constant prefactor has been adjusted to scale the amplitude  as in Fig.\ref{totSumo}.

\section{the TE mode propagation in the superconducting phase}

\subsection{Why photons should propagate    in the superconducting phase, below the AH threshold}

\begin{figure}
      \includegraphics[height=64mm]{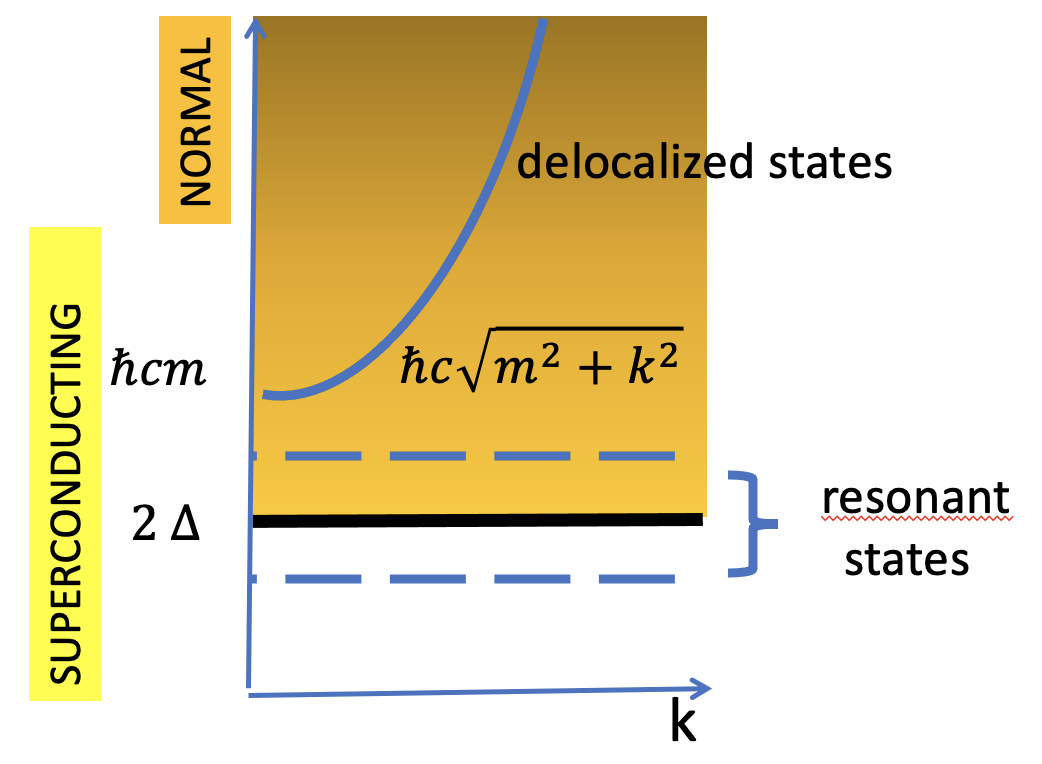}
\caption{ sketch of the energy dispersion  versus $k$  for $k_\parallel =0$. Two resonant states are sketched below the band of delocalized states of the massive   Anderson-Higgs modes, as  they merge in the gapped energy window.  $c$ is the photon velocity in the medium and $m = 2\pi / \lambda _L^\perp$, where $\lambda _L^\perp$ is the London penetration length in the $c$ direction for YBCO. At high energy the linear dispersion of the normal phase is recovered.   } 
   \label{gap}
\end{figure}

 In the superconducting phase, the Anderson-Higgs  (AH) mechanism makes the three e.m. modes massive, with mass $ \hbar c  m=2\pi  \hbar c  / \lambda _L^\perp $, where $ \lambda _L$ is the London penetration length of the field components into the sample.   Here $c$ is the photon velocity in the medium. Propagation only occurs  at energy $ >   \hbar c  m$ with the dispersion  $ \omega _{\vec{k}}/c = \sqrt{m^2 + k_\perp^2+ k_\parallel^2}$. The two transverse massive modes are similar to the  TM mode of the normal phase  at the surfaces, but they decay  in the interior of the material. They both couple to the MS surface excitation mode.   In a macroscopic approach (i.e. based on a model for $\epsilon (\omega)$), the TE mode does not couple to surface plasma modes in the ideal normal metal film, at least within first order perturbation theory. This is the reason why it is usually assumed that  the TE photon contribution to the Casimir energy is quite scarce in the normal phase. In the superconducting phase,   the  longitudinal massive photon mode  can be assimilated to a TE mode, because of the non vanishing  $B_z$ component.    As $E_z =0$,  we are confident that no current is injected in the superconductor, a crucial requirement  at low frequencies. However, being massive, the longitudinal mode  should not propagate across the slab if it is relatively thick.  Close to the transition temperature, the penetration length $\lambda _L^\perp$ is quite long, ($\sim 2.6 \: \mu m$ at $ T \sim 86.5 ^\circ K$). Hence we can expect that the length of the sample $ a \lesssim \lambda _L^\perp $.  Away from the transition temperature, the AH mass is rather  large and states with energy above it are not expected to contribute much  differently between the normal and superconducting phase. In fact, as in the case of the CG  TM mode,  the large enhancement of  qp excitations in the density of states of  the superconducting phase  at the pair breaking  energy  suggests that   TE photon tunneling can be  assisted by  virtual excitations with qp's production in the $a-b$ planes.   Indeed, the pair breaking energy is much lower than $ \hbar c  m$  in HTS  (see Fig.\ref{gap}).  However,  question arises if the longitudinal mode takes advantage of    photon resonances  at energy below  $ 2\Delta < \hbar c  m$, to propagate across the sample.  Resonances can be induced by virtual coupling with the in-plane superconductivity, originating from virtual excitations  with broken pairs bound of the  $a-b$ planes.  The answer is positive. The search for these resonances is the content of the  subsections III.C,D. They characterize the superconducting phase and are expected to give an appreciable  contribution to the  Casimir energy difference. 
 
 As discussed in the Introduction, on the one hand we cannot account for the microscopic structure of the array of $CuO$ planes in the lattice. The scale of $k_{\perp}$  for photons interacting with the  planes  in the lattice is of the order of the inverse of  the  lattice spacing, which, in YBCO  is $  a \sim 3\: nm $. On the other hand,  a photon in the micro-infrared frequency range  can only see  a mediated structure of  cells. We will adopt a  scattering approach  for a model structure and we will  show that  virtual pair breaking processes in interaction  with  the photon field   allows for  resonant  longitudinal states in the AH gap. The  TE   photon  modes  are  well defined and long lived as long as they are  located in energy  below the $2\Delta $  threshold and  contribute to the Casimir energy.  In our model we assume no space dependence in the $a-b$ plane,for simplicity,   which corresponds to $ k_\parallel  =0$ and we will drop the label $\perp$ in the following. 

We now describe the model interaction  in some detail.  We  assume a bulk HTc  material with planar boundary surfaces and consider scattering in the $\hat z$ direction, orthogonal to the surface,  with  $\hat z$ parallel to the $c$ axis for simplicity. This implies that the surfaces exposed to the  impinging radiation are  flat $a-b (Cu O)$ planes. The vacuum radiation  of energy $\hbar \omega_{\vec{k}} /2$ is characterized by a  component of the wavevector $k$ orthogonal to the planes and a transverse component $k_\parallel$, parallel to the planes.   
A TE photon of infrared frequency, with  a wavevector component  $\kp$ in the surface plane,  can break a number $N$ of pairs.  $N$ is of the order of $10^4\div 10^6$ for microwave photons. However, as the film is  macroscopic and superconducting, it  does  not  conserve the pair number  anyhow.  Let  $\epsilon = -\Delta $  be the binding energy of a pair. We consider as Ground State (GS) of the system, the state of the superconducting plane of  energy $N\epsilon $ in which   $N$ pairs are  unbroken \cite{anderson} and there is no real photon and we denote it  by 
 $| 0,\uparrow \rangle $. On the other hand, $|1, \downarrow \rangle $ is the excited  state  in which   $N$ pairs are broken and a real photon  is present, trapped in the film. Let us assume that a potential matrix element  $\Omega $ couples these two states and the Hamiltonian applied to these states, with $k_\parallel \sim 0 $, reads:  
 \bea
H_N  \left (  \begin{array}{c}  \left |1,\downarrow \right \rangle \\    \left |0,\uparrow \right \rangle \end{array} \right ) = \left (  \begin{array}{cc} -N\epsilon + \hbar \omega _k  &\Omega \\   \Omega  &  N\epsilon \end{array}\right ) \left (  \begin{array}{c}  \left |1,\downarrow \right \rangle \\   \left |0,\uparrow \right \rangle \end{array} \right )\:\:\:\:\label{amo}\\
 = \frac{\hbar \omega _k}{2} + \met \left (  \begin{array}{cc} - 2 N\epsilon + \hbar \omega _k  &\Omega \\  \Omega  &  2N\epsilon - \hbar \omega _\kp\end{array}\right ) \left (  \begin{array}{c}  \left |1,\downarrow \right \rangle \\  \left |0,\uparrow \right \rangle \end{array} \right ). \nonumber
\enea
The eigenvalues are: $ {\cal{E}}_\pm   =  \frac{\hbar \omega _k}{2} \pm  \met    \sqrt{ \delta ^2+\Omega ^2 }$, with $\delta = 2 \:  N\epsilon - \hbar\omega _k$: 
\bea
 {\cal{E}}_-   \approx   \frac{\hbar \omega _k}{2} -  \met  | \delta |  \sqrt{ 1+\frac{\Omega ^2}{\delta ^2}  } = 
+ N\epsilon  -\frac{1}{4} \frac{\Omega ^2}{|\delta |}, \nonumber\\
 {\cal{E}}_+   \approx   \frac{\hbar \omega _k}{2} +  \met  | \delta |  \sqrt{ 1+\frac{\Omega ^2}{\delta ^2}  } = \hbar \omega _\kp 
 - N\epsilon  +\frac{1}{4} \frac{\Omega ^2}{|\delta |}.  \nonumber
 \enea 
 The state corresponding to  $ {\cal{E}}_- $ corresponds to a state 
 \beq 
 | -  \rangle = \cos \theta  \: |0,  \uparrow\rangle +  \sin  \theta  \: |1,\downarrow\rangle 
  \label{GSstate}
 \eneq
   with $\theta$ close 1   and is the GS of the system, while  the excited state   corresponding  to energy  $ {\cal{E}}_+ $ is
    \beq
   | +\rangle = -\sin \theta  \: |0, \uparrow\rangle + \cos  \theta  \: |1, \downarrow\rangle .
   \label{eccstate}
   \eneq  
    Higher excited states are disregarded.
    
   In a scattering approach the interaction is localized in the film, while  the incoming photon  and the superconductor, very far from the scattering area and  in the vacuum, are  in the  uncoupled  state  $  |\Psi_0\rangle =  \left | 0  \right \rangle\!\! \left  |\uparrow \right \rangle $.The  pair number is not conserved, so that  we can assume that  the energy   $ {\cal{E}}_- $ is equal to the energy of the state $|\Psi _0\rangle$, in which the incoming photon and the superconductor are uncoupled, neglecting second order contributions  to the energy in the coupling $\Omega$. 
  We discuss the  zero temperature case and the channel  of energy $ {\cal{E}}_+  $  is  closed. 
 
\subsection{Scattering approach to the longitudinal mode propagation}

\begin{figure}
\includegraphics[height=50mm]{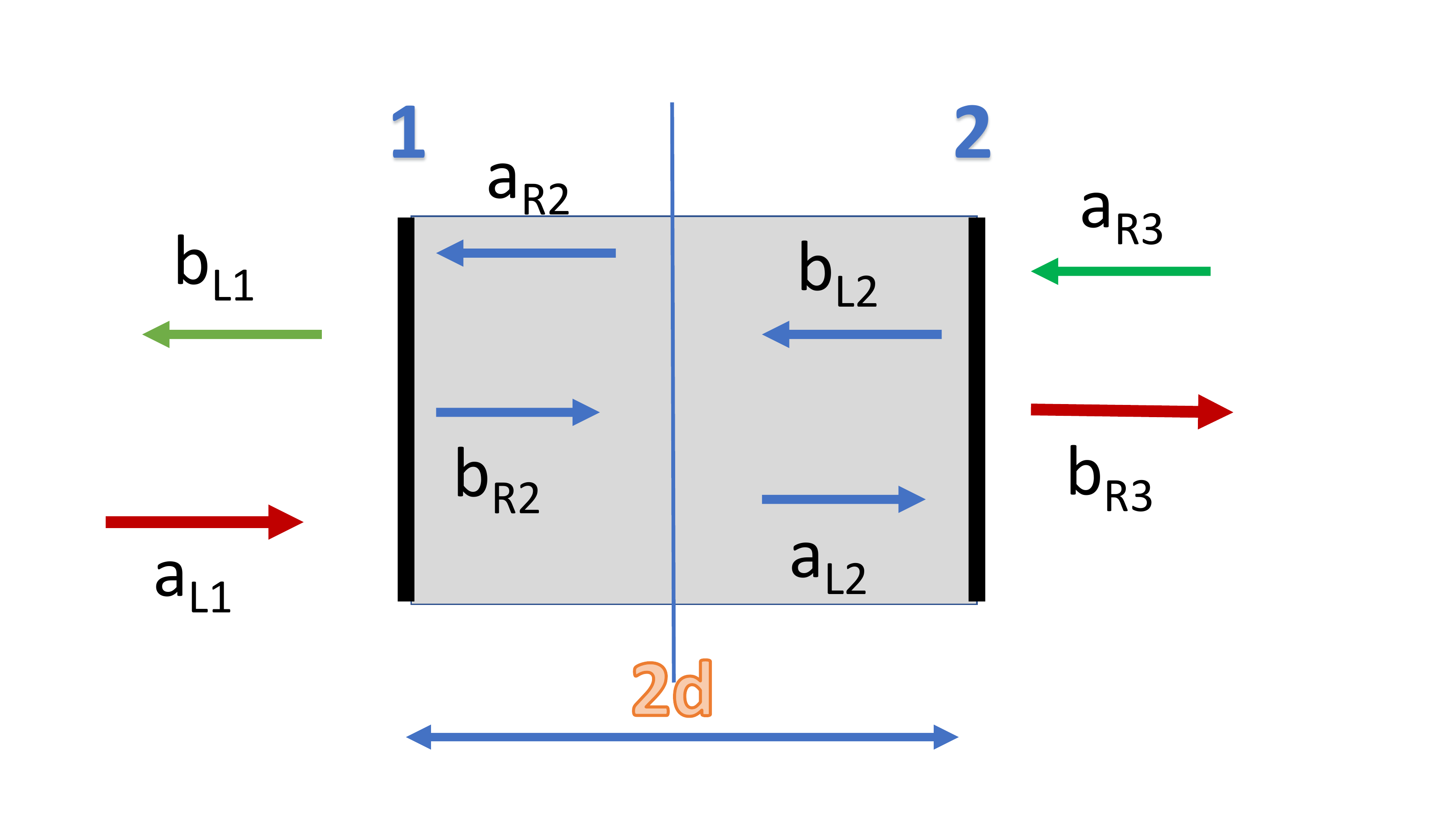}
\caption{ sketch of  the model film ({\sl grey zone}) with two $\delta-$functions ({\sl black thick lines at the boundaries}) at a  distance $ 2 d$.  $a$ amplitudes  refer to  incoming channels, while $b$ amplitudes refer to outgoing channels.  } 
   \label{cell}
\end{figure}

  We first discuss the scattering  of a virtual photon from the vacuum into  the AH modes  inside the superconductor, at energy above the  AH mass threshold $\hbar c m$.  Being the AH modes longitudinal,  it can be  matched with the TE mode impinging on the superconductor surface. 
 The wavefunction of the  photon of wavevector $\vec{k}$ is delocalized  everywhere in the  space at the  left ($L$) hand side of  the metal  chunk and it is  scattered and transmitted  to the  right ($R$) hand side of it. 
 To characterize the scattering of a photon on the superconductor, at least  in the limit of $k_\parallel \to 0$,  the simplest  scattering  approach will be adopted, with   two  $\delta -$potentials  at distance $2 d$  to  mimic the matter-radiation  model  interaction at the   two planar surfaces of the   superconducting  film (see Fig.\ref{cell}).
 To keep some analogy between the scattering approach and  the original geometry, we have to to include also the very left space region and very right side one,  as in Fig.\ref{cell}.  The total length of the scattering region, symmetric with respect to the origin,  is $a= 4d$.


 
      


            To  show how  the boundary conditions  for the electric field  are set  at the  film surface, we  first consider just one planar surface  interaction at $z=0$ in free space. 

The "incoming" state is $|\Psi _0\rangle =  \left | 0  \right \rangle\!\! \left  |\uparrow \right \rangle  $. We denote just by $k$ the component   $k_z$ orthogonal to the surface plane and  we make explicit  the label for the parallel component of the $k$ vector, $\kp$. The wavefunctions   $\psi _{L,R}$, defined outside the scattering region at $z=0$ are:  
\bea
 \psi _L &=& e^{ikz} \: |0, \kp\rangle + r \: e^{-ikz} \: |0, \kp\rangle   + s  \: e^{\kappa _Lz} \: |1, \kp\rangle \nonumber\\
  \psi _R &=& t\:  e^{ikz} \: |0, \kp\rangle  + \tau   \: e^{-\kappa _Rz} \: |1, \kp\rangle .  
  \label{eqpart}
 \enea
 $r $ and $s$ ($t  $ and $\tau $) are reflection (transmission) amplitudes  for the two channels of transverse wavevector $\kp$.  $ |0, \kp\rangle $ and $ |1, \kp\rangle $ are photon states.
The channel $ |1, \kp\rangle $ is assumed to be closed, so that  $\kappa_{L/R} $ are real parameters depending on the incoming energy and on $k_\parallel $, which  is assumed to be conserved.

The superposition of the states  $\Psi _0$,  with    $ \Psi , \ppsi$ defined above,  due to the interaction, provides the  field  wavefunction at fixed  $k_\parallel $, as a function of energy and $k$ orthogonal to the plane.  
 In the case of the $\delta-$function potential, the matching conditions require continuity of the wavefunction at the scattering plane,  $z=0$, and a jump of the space derivative there: 
 \bea
 |  \psi _L (z=0)  \rangle  &=&\left .[  | \chi_-  \rangle + \beta  | + \rangle ]\right |_{z=0} =  | \psi _R (z=0)   \rangle 
  \label{eqL}\\
    \left .  \frac{ d\left |\psi _L \right \rangle}{dz} \right |_{z=0^-} &-&  \left .  \frac{ d\left |\psi _R \right \rangle }{dz} \right |_{z=0^+}  = g\: V\: \left . \left \{ | \chi_- \rangle  + \beta  | +\rangle \right \}\right |_{z=0},  \nonumber\\
        \label{dev}
    \enea 
where  $ |  \chi _- \rangle =    | \Psi _{0} \rangle + \alpha  |-  \rangle$  ($  | \pm \rangle $ have been defined in Eq.s(\ref{GSstate},\ref{eccstate}) and we assume $ |\Psi _{0}\rangle $ and $|-\rangle  $ to have the same energy) and  $\alpha,\beta$ are complex numbers.

  Tracing away the state of the condensate in the plane, Eq.s(\ref{eqL},\ref{dev})  should be projected onto
     $      | 0,\kp \rangle\left [ |\uparrow\rangle + |\downarrow\rangle \right ] $ and  $   | 1,\kp \rangle\left [  | \uparrow\rangle+ |\downarrow\rangle \right ]$,   to derive the dependence of $\alpha,\beta $ on $ \kappa_L,\kappa_R, k$  as  reported in Appendix B.

 At maximum superposition,$- \sin\theta = \cos\theta = \frac{1}{\sqrt{2}}$, is: 
  \bea
 {\alpha}  +{\beta} =  \frac{ g\: V }{2  {\kappa}_+ }\left [1 + \frac{ {\kappa}_+}{2 \: i\: k }\right ] , \nonumber\\
    {\alpha}  -{\beta} =   \frac{ g\: V }{2  {\kappa}_+ }\left [1 - \frac{  {\kappa}_+}{2 \: i\: k }\right ]
    \label{alpbet}
   \enea
  ($\kappa_+ = \kappa_L+\kappa_R$). Unitarity of the S-matrix fixes the ratio $ \frac{gV}{2\kappa _+}$:
    \bea
  \frac{gV}{2\kappa _+}=  -i\:  \frac{ {\kappa}_+}{2 \: k } \: \frac{1}{1+  \left (\frac{ {\kappa}_+}{2 \: | k  | } \right )^2}.
  \label{pot}
  \enea
  By  taking  the inverse decay length corresponding to the Meissner effect in the superconductor, $\kappa _+ /2 = m$ all parameters are fixed, except a mixing angle $\eta$, so that  the 
   S-matrix for one single $\delta-$ barrier is ( $x=\frac{\kappa _+ }{2k}$) : 
   \begin{widetext}
     \bea
    S =  \left (  \begin{array}{cccc} r  & 0 & t & 0 \\ 0 & s & 0 &  \tau \\t & 0 & r' & 0 \\ 0 & \tau & 0 & s' \end{array}\right ) =
     \left (  \begin{array}{cccc} -\frac{i\: x\: u}{1 + i\: x}  & 0 &  \frac{\sqrt{1+ x^2 v^2}}{1 + i\: x} & 0 \\0 & -\frac{i\: x\:v}{1 - i\: x}  & 0 &   \frac{\sqrt{1+ x^2 u^2}}{1 - i\: x}  \\ \frac{\sqrt{1+ x^2 v^2}}{1 + i\: x}  & 0 &  -\frac{i\: x\: u}{1 + i\: x}  & 0 \\ 0 &   \frac{\sqrt{1+ x^2 u^2}}{1 - i\: x}  & 0 &  -\frac{i\: x v}{1 - i\: x}  \end{array}\right ),
   \label{smatdS}
    \enea 
    \end{widetext}
    where $u= \cos \eta $ and $v=\sin \eta $. We have excluded direct 
 interaction between channel $|0,\kp \rangle $ and $|1,\kp \rangle$. Such an interaction would give  an output  amplitude in the $|1, \kp \rangle$  channel, which is an inelastic process, which would lead to  dissipation. At  $\theta =-\pi/4$ in Eq.s(\ref{GSstate},\ref{eccstate}), the parameter $\eta$  does  not  play any role, because,  being the channels independent, every dependance  on $\eta$ is washed out by unitarity.  The restriction $t'=t$ adopted here is allowed  in the search of bound states  provided time reversal holds. Eq.(\ref{smatdS}) extends the $S-$matrix for elastic scattering  with one single channel;
      \bea 
 \left (  \begin{array}{c} b_L   \\ b_R\end{array}\right )=  \left (  \begin{array}{cc} r & t' \\t & r'  \end{array}\right )
  \left (  \begin{array}{c} a_L \\ a_R  \end{array}\right ),  \label{smat}
    \enea 
 The wavefunction amplitudes  $a_{L/R}$ are  the {\sl in}-wavefunction amplitudes, while   $b_{L/R}$ are  the {\sl out}-wavefunction amplitudes for the  AH  mode.  In our case, each element is a $2\times 2$ matrix because it includes the channel label $0,1$, corresponding to photon states  $| 0,k\rangle $ and   $| 1,k\rangle $.  

Now we turn to the geometry of Fig.\ref{cell}, by using the following  procedure\cite{minutillo,guerout}. The $S-$matrices  $ S_{1,2}$  of each of the  $\delta-$functions are   translated by  $\pm d$, respectively with respect to the origin,  by means of an unitary matrix  $ \Lambda (\pm d) = diag \left [ e^{\pm i k d}, e^{\pm i k d} \right ]$, where $k$ is the  $k-$vector corresponding to the energy of the incoming photon.  Next, the  transfer matrices  corresponding to   $S_{1,2}$ are derived, defined as: 
 \bea
  \left (  \begin{array}{c}   b_{Ri+1}\\  a_{Ri+1}\end{array}\right )=  {M}_{i} \:    \left (  \begin{array}{c}   a_{Li}\\  b_{Li}\end{array}\right ). 
  \nonumber 
  \enea
   The chaining  $ M_2 * M_1$ corresponding  to  matrix multiplication provides:  
   \bea
   \left (  \begin{array}{c}   b_{R3}\\  a_{R3}\end{array}\right )= M_2 M_1 
   \:    \left (  \begin{array}{c}   a_{L1}\\  b_{L1}\end{array}\right ) \nonumber
   \enea
   Final step is to transform back the full transfer matrix to  give the  global scattering matrix $S'$,
   \bea
   \left (  \begin{array}{c}   b_{L1}\\  b_{R3}\end{array}\right )=  S'\:  
   \:    \left (  \begin{array}{c}   a_{L1}\\  a_{R3}\end{array}\right ),  \label{sprim}
   \enea
with the result:
 \begin{widetext}
\bea
 S' \equiv  \left (  \begin{array}{cc}  e^{-i\:k a}  \: s_{11}  & e^{- \:2\: i\:  k (a+d) }\:  U_\delta \: s_{12}  \\ e^{ 2\:i\: k d}\:  s_{21}  & e^{-i\: k a} \: U_\delta  s_{22} \end{array}\right ) \equiv   \left (  \begin{array}{cc} S_{11}  & S_{12}  \\ S_{21}  & S_{22} \end{array}\right ) \hspace{2cm} \nonumber\\
 {\rm where} \hspace{12cm}\nonumber\\
 s_{11} =  r_1 +  {t_1'}\left ( 1-   t_2\:  {r'_1 }\right )^{-1} t_2 t_1 \label{matamat}\\
s_{12} = {t_1'}\left ( 1-   t_2\:  {r'_1 }\right )^{-1}  {r_2'}\: e^{-2i k d} \nonumber\\
s_{21} = e^{2i k d} {r_2} \left [ 1 -   r'_1 t_2 \right ]^{-1}  {t_1 }
  \nonumber\\
  s_{22}  =t'_2 +e^{2i k d}  r_2\:  {r_1'} \left ( 1-   t_2\:  {r'_1 }\right )^{-1}  {r_2'}\: e^{-2i k d} \nonumber\\
   U_\delta =  e^{2\: i\: kd}\:  e^{2\: i\: k \:(a+ d)}\:  s_{11}  s_{22}^{-1\dagger} 
  s_{12}^{\dagger } s_{12}^{-1} \equiv e^{2\: i\: ka}\:  \left (  \begin{array}{cc} e^{2\:i\: \delta_1 } & 0  \\0 &e^{2\:i\:\delta_2 }   \end{array}\right ). 
\hspace{1cm}\nonumber
  \enea
  \end{widetext}
   $ r_i$ and $t_i$ are the $2\times 2$ matrices defined in Eq.(\ref{smatdS}). Translation by  $\Lambda(\pm d))$ implies that the matrices $r_2, r'_2$  acquire a phase  $ e^{\pm 2\: i\: kd}\: $ with respect to $r_1, r'_1$.   $\delta_{1,2}$ are  the phase shifts of the two channels within the cell due to the scattering. 
 This is the result of  
 Gu\'erout et al.\cite{guerout}. Note a small difference in the ordering in  $S'_{11}$.
Numerically, our scattering matrix  is found to be unitary.
 
   The $S-$ matrix is numerically found to be unitary. Besides, as can be checked numerically:
   \bea
   S_{11}  S_{22}^{-1\dagger} 
  S_{12}^{\dagger } S_{12}^{-1} =1 \:\:\:\:\:\:  and \:\:\:\:\:\:  S_{12} =-S_{21}.
 \label{unit}
  \enea 
  From the definitions of $S_{ij} $ and $s_{ij} $ and the last equality we get:
\bea 
U_\delta \equiv - e^{4\: i\: k d }  s_{21} \: s_{12}^{-1} 
\label{shif}
\enea
Note that, in the case of elastic scattering with a single channel, if we put: 
 \bea
  { S'}\:   =  \left (  \begin{array}{cc} r& t^*  \:  e^{- 2 i\delta}  \\  t &  -r^*  \:  e^{- 2 i\delta}  \end{array}\right ),
  \enea
the condition $   S_{11}  S_{22}^{-1\dagger}  S_{12}^{\dagger } S_{12}^{-1} =1$  provides  $ r \left ( -r^{-1}  \:  e^{- 2 i\delta} \right )\: t \:  \:  e^{ 2 i\delta}   \: {t^*}^{-1}   \:  e^{ 2 i\delta} =1 $, that is   $  \:  e^{- 2 i\delta}  =- t/ t^*$, as expected.

   \subsection{TE  resonant contribution to the Casimir energy for the superconducting  phase}
\begin{figure}
\includegraphics[height=60mm]{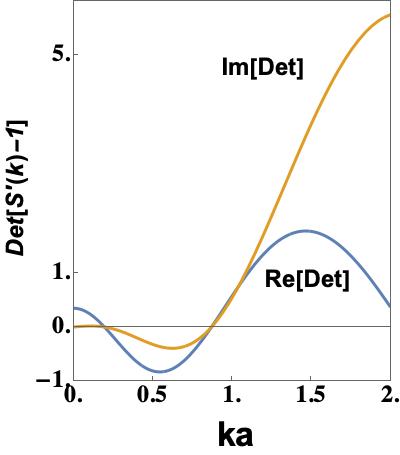}
\caption{ Real and Imaginary part of the  $Det \left \{S'\left [ \overline{ k} \: \right ] - {\bf 1} \right \}$ vs. $\overline{k} = k/m $ for $a= 0.378 \lambda _L$ and $k_\parallel =0$.  In these units the AH threshold is marked by $\overline{k}=1$. Two resonant  modes  at low energy correspond to the zeros of $Det \left \{S'\left [\: \overline{k}\: \right ] - {\bf 1} \right \}$. }
   \label{cro}
\end{figure}

\begin{figure}
      \includegraphics[height=60mm]{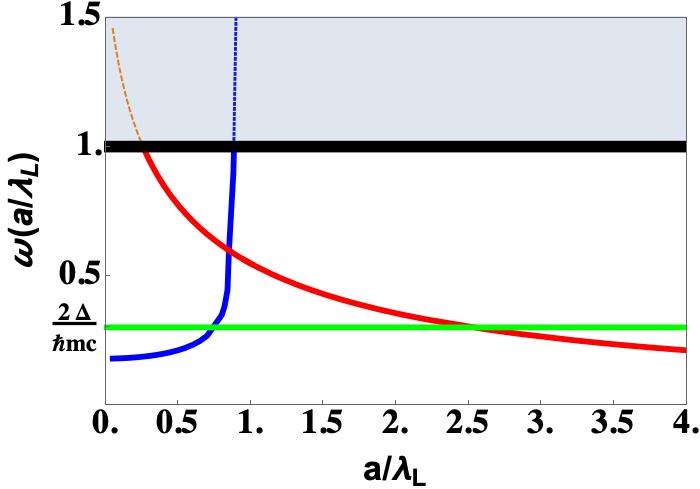}
\caption{ energy $\omega $ of the resonant longitudinal  states  (normal incidence), below the massive   Anderson-Higgs  propagating  modes ({\sl{grey area above the black horizontal  line}})  inside the superconductor versus $a/\lambda _L $. $a=4 d$ is the length of the model sample  in the direction orthogonal to the $Cu O$ planes. Energy, in the limit $k_\parallel =0$, is in units $\hbar c m$, where  $c$ is photon velocity in the medium and  $m= 2\pi /\lambda _L$.  The red curve is the mode  odd for $L \leftrightarrow R$  inversion, while the {\sl blue curve}  is even and has low energy only for $a < \lambda_L$.  The {\sl green dashed line} marks the threshold for pair breaking excitations, at  $2\Delta$. The vertical axis is not in scale.}  
   \label{randa}
\end{figure}

  In the case of the superconducting phase,  extended propagating states below the  Meissner, AH  threshold are not allowed.  However,  analysis of the $S-$ matrix of Eq.(\ref{sprim},\ref{matamat}) shows that there can be one or more resonant  states  propagating across  the superconductor, below the Meissner threshold, as sketched in Fig.\ref{gap}.  Their signatures are by the zeros of the determinant  $Det \left \{S'\left [\: k\: \right ] - {\bf 1} \right \}$.  In Fig.\ref{cro} we report a plot of the real and imaginary part of the determinant. 
 The zeros 
 appear  at   energies  $\overline{k}_1\equiv  \omega_{\overline{k}_1} / (\hbar c m ) =0.2$ and $\overline{k}_2  =0.9$ (in dimensionless unities),  for a length of the sample $ a= 0.378 \lambda _L$. Here $k_\parallel= 0$ (normal incidence), for simplicity. 
Fig.\ref{randa} shows the  energy trend of these two states with increasing length of the sample.  The {\sl horizontal black line} is the AH threshold and the propagation modes are fully delocalized above this energy.  The {\sl green dashed } line marks energy $2\Delta$, one tenth of  $\hbar c m $ (the $y-$axis is not in scale). In the energy interval  $(2\Delta, \hbar c m)$ single qp's are produced by pair breaking and the modes acquire a finite lifetime.   When the two  modes   are in the energy window $ < 2\Delta $  in which a continuum of propagating modes is forbidden, they act as resonances in the propagation of the field.  For  $ a < \lambda_L$, which corresponds to  full  penetration of the radiation inside the superconductor, the {\sl blue curve} resonant mode is even with respect to the inversion center of the sample and is lower in energy. However sustaining radiation inside the superconductor costs much energy  when $a \approx  \lambda_L$  and the even mode increases sharp with a very short lifetime (only the real part of the energy appears in the plot).  For   $ a > \lambda_L$ the odd mode  ({\sl red curve}) becomes lower in energy because it allows for small field amplitude with a node a node   inside the superconductor.   We renounce  to qualify the field amplitude  within the sample but we infer the parity of the modes from the parity of the phase shift across the sample when $ k \leftrightarrow -k$. We expect that non normal incidence ($ k_\parallel \neq 0$) would mix the two modes particularly  at intermediate lengths $a \sim \lambda _L^\perp$,  by opening  a gap a the crossing of the curves. 

\section{total Casimir energy  and energy difference}
The total Casimir energy  in the normal phase has been discussed in Section III.C . Here we present our estimate for the  total Casimir energy in the superconducting  phase and the difference between the two. 
   \subsection{total Casimir energy in the superconducting  phase}

 The total Casimir energy in the superconducting phase  does not include propagating states below the  Meissner, AH  threshold $\hbar c m$, except for the TE resonances. In our estimate we assume that the contribution coming from  energies above the  Meissner  threshold  and from the  qp's  in the  energy window $(2\Delta, \hbar c m)$  is roughly cancelled by a  corresponding contribution in the normal phase, when we eventually take   the difference. In fact,  single qp delocalized states are present both in the superconducting and in the normal phase. The contribution to the total Casimir energy difference due to the marked change in the density of states close to the $2 \Delta $ threshold,  between the two phases,   is discussed in Appendix C. The $2\Delta $ gap threshold  induces a sizeable change of the dielectric function, as discussed in Section II  with important changes in the photon propagation at  that  energy range. However,  if we are at temperatures rather away from  $T_c$, we can expect that the weight of this contribution, is scarce  for microwave photons and we will ignore it.  It  is considered to be small and is neglected. There are no propagating states at energy below the gap threshold  $2\Delta$, so that   the only  contributions  to the Casimir energy which we consider for the superconducting phase arise from  the TM plasma mode and the TE resonance (just one at the chosen lengths of the sample).  
 
  The  symmetric  TM  mode,  even with respect to $ L \leftrightarrow R $ inversion symmetry, appears in Fig.\ref{mooij}. It is  linearly dispersed in $k_\parallel$ at small $k_\parallel$  values, while   is dispersed as $\sqrt{k_\parallel}$ at larger $k_\parallel$. We follow the same steps as in  Eq.(\ref{noret}) to subtract the $ a \to \infty$ term and leave just the $a$ dependent contribution. Using Eq.(\ref{disdis}) and   cutting  the  $k_\parallel  a/2 = s$ integration at  $\overline{s} =2\Delta \sqrt{2 \epsilon _s} / 	\omega _0$, the contribution to the Casimir energy  of the TM mode is approximately:
  \bea
  E^{TM}_{Sup} = \met \sum _{k_\parallel} \omega _{k_\parallel} = -\met  \left (\frac{2  L}{2 \pi a}\right )^2 \:  2\pi  \int _0^{\overline{s}} \frac{s^2 }{2 } ds\:   \frac{\partial \omega (s)}{\partial s} \nonumber\\
  \approx  -  \left (\frac{ 2 L}{ a}\right )^2  \frac{\omega _0}{2\pi } \: 0.037, \hspace{1cm}
  \label{TMsup}
  \enea
  in analogy with Eq.(\ref{noret}). Here  $\omega _0 \lesssim 2\Delta $. 
  Based on the fact that the MS mode has a $\sqrt{k_\parallel}$ dependence on $k_\parallel$,  we  estimate  the integral in Eq.(\ref{TMsup})  by assuming  $\omega _{ps} \sim \omega _p$ and $ \omega _0 \sim 2\Delta /2\pi$. The result is plotted in Fig.\ref{risot}  ({\sl blue curve}).
  
   The piling up of qup excitations near the gap threshold allows for an odd mode  (which is  the  'neutral' ASPM) at those energies but only at larger $k_\parallel$vectors\cite{mooij}. Their influence is detected, according to our model, in the resonances that a TE photon propagating mode  can encounter at low energy according to Fig.\ref{randa}. This feature is absent in the  normal metal phase.   The contribution of the TE mode  to the Casimir energy is negative  for  $ \lambda_L ^\perp > a$,  i.e. when  the working temperature is not far from $T_c$.  Here we give an estimate of the TE resonance   for  $a= 0.378 \lambda _L^\perp $ (see Fig.\ref{cro} and Section IV.C). 
   The energy of the  resonance disappears  for  $a\to \infty$, so that we do not have to subtract any  $a-$independent limiting contribution. The  energy of the resonance is given by the  zero of the determinant $Det \left \{S'\left [\: \overline{k}\: \right ] - {\bf 1} \right \}$  (with $S'$ given by Eq.(\ref{matamat}) ) and takes the value $0.2 \hbar c m $ when $k_\parallel=0$. However   its $k_\parallel$ dependence is weak, except for the fact that  direct tunnelling across the resonance  does not contribute to the Casimir energy.  Therefore, we add an angular dependence $(1-\cos \theta)$ in the integration over $k_\parallel$  and   approximate the contribution as follows, with $s= \frac{ k_\parallel a}{2}  \in \left ( 0, s^{max} \right )$, where   $s^{max} \sim    \frac{2\Delta}{\omega_0}  \sqrt{ \frac{ \epsilon _s}{\pi}} $: 
   \bea
   E^{TE}_{Sup} \approx   -\met  \left (\frac{2  L}{2 \pi a}\right )^2  0.2\:  \hbar c m\:   \int_{-\frac{\pi}{2}}^ {\frac{\pi}{2}}\!\!\!\!\!\ d\theta (1-\cos\theta )  \int _0^{s^{max} }\!\!\!\!\!\! \!\!\!\!\!\! s\:  ds \nonumber\\
   = -\met  \left (\frac{2  L}{2 \pi a}\right )^2  0.2\:  \hbar c m\:  \left ( \frac{\pi}{2} -1 \right )  \left ( \frac{2\Delta}{\omega_0}  \sqrt{ \frac{ \epsilon _s}{\pi}}  \right )^{4}\!\!\!\!\!. \hspace{0.4cm}
   \label{TEsup}
   \enea
   The dispersion in energy vs. linear size of the sample appears in Fig.\ref{risot} ({\sl orange curve}). Its weight in the  density of states is rather small and this implies that   it gives a little  contribution to the Casimir energy. In particular, the contribution changes sign at $\lambda _L^\perp \sim a $ (see Fig.\ref{randa}), but it is anyhow vanishingly small for $a > \lambda _L^\perp $. 
    
   In our model,  the TE resonances arise from bound states that are split off  the delocalized AH band with threshold $\hbar c m k=1$ in our units. In the superconducting phase there is a continuum of electronic qp states of energy above the pair breaking threshold energy $ 2 \Delta$. They could contribute to the transfer of photons across the sample, so that we can assume that there is  a continuum of photonic states corresponding to their energy.  We comment on these delocalized photonic states  here in the following.  Our model system  acts as a $1-d$ potential well of length  $a$ which can bound states. 
 As a function of energy $\omega$ the change of the density of states   due to the scattering,  derived from the Green's functions defined by  $ G = G_0 + G_0 \: t \: G_0 $ ($t$ is the $t-$matrix   defined in Appendix A) is  given by: 
\bea
\Delta \nu (\omega )  = - \frac{1}{\pi} \: \Im m \: Tr \left \{ G^R-G_0^R\right \}  \nonumber\\
 =   \frac{1}{2\pi  i } \frac{d}{d\omega }\: \ln\:  Det\: S'(\omega ) 
  \label{nuchange}
 \enea
(the label $R$ denotes 'retarded Green's function'). As  $\left |  Det\: S'(\omega )  \right | =1 $,  $\Delta \nu (\omega )  =   \frac{1}{\pi  } \sum _j \frac{d}{d\omega }\: \delta _j(\omega )$, where  $ \delta _j $ are the phase shifts of the two channels ($j = 1,2$). 
 \begin{figure}
 \centering
\def\big{\includegraphics[height=4.8cm]{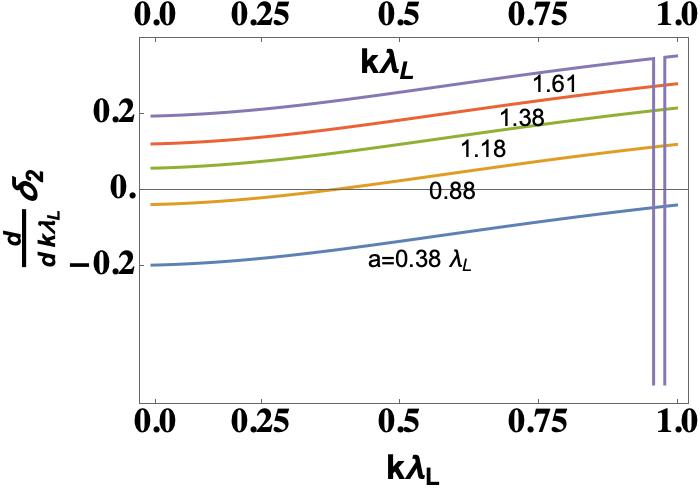}}
\def\little{\includegraphics[height=2.7cm]{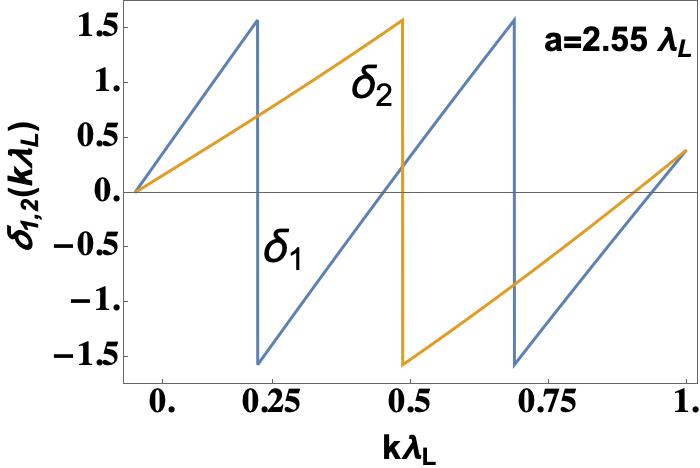}}
\def\stackalignment{l}
\topinset{\little}{\big}{+80pt}{+95pt}
\caption{Derivative of the phase shift $d \delta_{2} /d k $ vs energy $k$ (in dimensionless units) for  $a = 0.38, 0.88,1.18, 1.38,1.61$ ({\sl from bottom to top}).  At $a\approx 1.6 $ a bound state  splits off  the bottom of channel 2, $k=1$ and contributes with a $\delta -$like peak to the derivative ({\sl exaggerated in the picture}). {\bf \sl inset}: Phase shift $\delta_1$ ({\sl blue}) and  $\delta_2$ ({\sl orange})  for $a = 2.55$.  The $\pi -$ jumps  in  $\delta_{1,2}$  mark  two bound states in channel 1  and one bound state in channel 2. } 
\label{shifts}
\end{figure}

 The matrix $S'$ can be set in a  block form,  diagonal in the channel label $j$. The contributions of Eq.(\ref{nuchange}) coming from the phase shifts  $ \delta_{1,2} $ should be included in our estimate of the Casimir energy  for the longitudinal mode and  compared with the corresponding ones of the TE mode of the normal phase.  In particular channel 1, would refer to processes which occur both in the normal metal phase and in the superconducting phase. As for channel 2, according to our model, its influence   is only limited to the superconducting phase and mimics processes in which propagation includes Cooper  pair  breaking events, close to energy $2\Delta$.  
  A similar contribution was presented in the macroscopic approach for the TM modes  in Section  II.B. We argued there that pair breaking processes make the largest difference, but  can be assumed to have little role  at our  much lower incoming photon energies, except for virtual excitation. We are not including these contributions that had been already discarded in the case of the TM modes. 
 
In Fig.\ref{shifts} we have plotted the derivative of the phase shift $d \delta_{2} /d k $ vs energy $k$ (in dimensionless units) for various lengths of the sample  in units of $\lambda_L^\perp$.  A sharp drop for $k\lesssim 1$ in the curve for $a \lesssim 1.6$ marks the splitting of a bound state related to channel 2 from the bottom of the AH energy dispersion. The bound state appears as a $\pi -$jump in the phase shift $\delta _2(k)$.  Bound states appear as  $\pi -$jumps  in both  channels,  as shown in the  inset of Fig.\ref{shifts},  where the phase shifts   $\delta_{1,2}$  are plotted vs. $k$, for $a = 2.55$. In fact the potential formed by the two $\delta-$functions acts  as an attractive  potential well for the photons. It follows that    bound states are split from the bottom of the AH energy dispersion and move to lower energy with increasing  distance between the $\delta-$ peaks.  At given coupling strength, the threshold  thickness of the sample for the appearance of a bound state splitted  off channel 2 is $a \lesssim 1.6$.

\subsection{Casimir energy difference  $ \delta{\cal{E}} =  E_{Sup}-  E_{Nor}$ }
\begin{figure}[h]
\begin{center}
\includegraphics[scale=0.4]{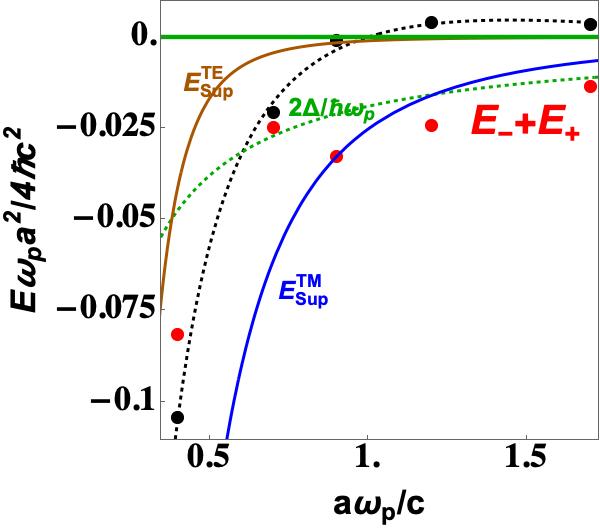}
\caption{ Various contributions to the Casimir energy difference (per unit surface)  vs  linear size of the sample. The {\sl red dot} are the contribution given by the plasma excitations in the normal phase ({\sl  from Fig.\ref{totSumo}}), labeled as $(E_++E_-)$.   $E^{TM}_{Sup} $ is the TM plasma mode in the superconducting phase    ({\sl blue curve , from  Eq.(\ref{TMsup})}). $E^{TE}_{Sup} $ is the TE plasma mode in the superconducting phase induced by coupling with pair breaking processes    ({\sl orange curve , from  Eq.(\ref{TEsup})}).  The difference  $ \delta{\cal{E}} =  E^{TM}_{Sup} + E^{TE}_{Sup} - (E_++E_-)$ is marked by the {\sl black dots} (the {\sl  dashed black curve} is a guide to the eye). The difference $ \Delta{\cal{E}} $ becomes positive at larger sizes but vanishes for size going to infinity. The threshold for pair breaking processes at various sizes is marked by the {\sl  dashed green  curve}.   } 
\label{risot}
\end{center}
\end{figure}
Fig.\ref{risot}  summarizes our estimates of the contributions to the Casimir energy  per unit area for a sample of linear size $a$. The {\sl black dots} are evaluations of the energy difference  $ \delta{\cal{E}} =  E_{Sup}-  E_{Nor}=  E^{TM}_{Sup} + E^{TE}_{Sup} - (E_++E_-)$ between the superconducting and the normal phase of the sample, at few $a$ values.  In our estimate only the contributions coming from the plasma excitations are included. In Section III.C, we have qualitatively estimated the contribution coming  from the  delocalized  photonic  states  in the normal phase as $ E^n_{Tot}-  (E_++E_-)$, but we have not included them. In the  energy range  $\omega > 2 \Delta$  they  are also present in the superconducting phase (although with a slightly different density of states except for energies in proximity of $2\Delta$), because  photonic transmission can be  assisted by the delocalized electronic  qp states at these energies and we can surmise that  these terms contribute roughly equally in the two phases. However, we have also neglected this contribution  for energies   $\omega <  2 \Delta$, which is present for  the normal phase  only, because,  as it appears in Fig.\ref{totSumo}, the energy difference $ E^n_{Tot}-  (E_++E_-)$ is rather small not only at small  sample sizes, but  even at larger sample sizes (we have plotted  also $2 \Delta$ in Fig.\ref{totSumo} ({\sl green dashed line}), which is devoted to the normal phase, for reference).  Besides $ E^n_{Tot}$ was estimated by means of  the scattering model of Section III.C and has been adapted, but is not homogeneous with the rest of the  calculation.  The brute approximation of neglecting $ E^n_{Tot}-  (E_++E_-)$  alltogether implies that larger linear sizes of the sample are not displayed in Fig.\ref{risot}. At those sizes, the two $\delta -$potential develops bound states also in channel 1,  as shown in the inset of  Fig.{shifts} and the scattering model becomes unreliable.  

 Inspection of the location of the {black dots} in  Fig.\ref{risot} vs linear size of the sample [$ (a \omega _p/c, \delta{\cal{E}})\approx  (0.4,-0.1),\: (0.7, -0.02),\:  (0.9, 0), $$\: (1.2, 4 \times 10^{-3}), \: (1.7, 3 \times 10^{-3})$],   shows that  the gain in Casimir energy  $ \delta{\cal{E}} $ when the sample undergoes the phase transition sharply depends on the linear size of the sample and can even become a loss when the size increases. This can be justified by noting that, in the normal phase, the absolute number of electronic qp states  increases with the size,   with an increase of the $| E^n_{Tot}| $ magnitude, while the $2\Delta $ gap in the superconducting phase  reduces chances for photon transmissions  assisted by qp's and hence, for contributions to Casimir energy gain. 
 
 A trade off between temperature and linear size of the sample is  also strictly required. On the one hand a temperature $T<<T_c$ implies that $\omega _{ps} \sim  \omega _{p} $, because the density of pairs exhausts the full electron density, and $ \delta{\cal{E}} $ would  increase.  But, in  the London theory, $c/\omega _{ps} \sim  \lambda _L^\perp$ and a shorter $ \lambda _L^\perp$ (for $T<<T_c$) implies that we move to larger   $a\omega _{p}/c$ values with a sharp reduction of  $ \delta{\cal{E}} $. On the other hand,  a temperature closer to $T_c$ would increase $ \lambda _L^\perp$ and  move to lower  values of  $a\omega _{p}/c$ of Fig.\ref{risot}, thus increasing  the gain in Casimir energy   $ \delta{\cal{E}} $, but $\omega _{ps}$ would become much smaller than  $ \omega _{p} $ and  the magnitude of  $ \delta{\cal{E}} $ is reduced. Besides, fluctuations would dramatically increase, especially in a  HTS, with a distructive role. In our derivation we have been choosing  $\omega _{ps} \sim  \omega _{p} $. 

Assuming a cubic sample, so that $L = c/ \omega _{p} = a$ and by choosing a reference value for  $\hbar \omega _{ps} \sim  \hbar \omega _{p}  \approx 40 \: meV$  we get   $ \delta{\cal{E}} \sim -1.4  \times  10^{-5} eV$. This is an optimistic  reference energy scale,  with more than an order of magnitude uncertainty. A better characterization of the result requires  the  choice for an appropriate temperature, which also depends on  estimates of the refraction index of the sample in the normal and superconducting phase in the range of  microwaves and of the plasma frequency in the two phases.

  \section{Summary and Conclusion}
 
 
 The Archimede project  is designed for measuring the effects of the gravitational field on a Casimir cavity by performing a weighing measurement of the vacuum fluctuation force on a rigid Casimir cavity\cite{avino,calloni,Allocca:2012kw}. This paper discusses the various contributions to the Casimir energy assuming that the "cavity" is just a metal bulk sample (a cube or slab) in vacuum. As a reference metal we take YBCO, which undergoes the superconducting phase transition at $T_c \sim 90\:  ^\circ \!K$.  The experiment will measure differences in weight between the superconducting and the normal phase by weighting at two different temperatures, above and below $T_c$. 
 
 A key point of the interpretation of the results of the experiment will be the estimate of the contribution of the Casimir energy to the total transition energy in the two phases and correspondingly, to the weight variation.
 
  It has been recently proposed that  the Casimir energy is a big part of the   "condensation energy", so that the driving mechanism for phase transition is the Casimir energy itself\cite{kempf}.
 
 Up to now, the Casimir force  has been measured in cavities of micron sizes\cite{sukenik} while the Casimir contribution to the transition energy for tens nanometers cavities has been theoretically and experimentally investigated within a previous experiment \cite{bimonte1, gig2, aladinExperiment1_2,Allocca:2012kw}, confirming
 the expected energy range for density of state changes in the photon field due to the presence of the cavity  corresponds to far infrared and microwaves.  At least in conventional superconductors where electron-phonon coupling is  considered as the pairing mechanism, the lattice parameter is the scale  at which  forces related to condensation energy act. Photons with a wavelength comparable to the lattice parameter have huge energy and  it is reasonable to expect that they propagate across the cavity with no harm whatsoever.

In the present work we limit ourselves to an estimate of the Casimir energy change  by comparing the zero point energy of the superconducting and the normal phase  in a macroscopic sample. 

There are various contributions to the zero point energy of the photon field.  Let us enumerate these contributions starting from the normal phase and continuing with the superconducting phase, afterwards.

In the normal phase  one  contribution arises from the continuum of  TM modes   propagating  across the sample in case the skin depth $\delta $  of the $E_z$ penetrating field is comparable with the linear size  of the metal slab (in direction $z$),  while  the continuum of  TE modes  should not contribute, except for tiny surface magnetization effects,  due to the reduced penetration of $ B_z$, in the case of a  paramagnetic material. Of course, propagation can be assisted by the continuum of electronic qp excitations in the sample via non elastic processes. The TM polarization contributes with plasma modes (charge excitations) localized at the surface in the normal phase. There are two plasma surface modes for the sample with two surfaces. In case of  a  $ z\leftrightarrow -z$  inversion symmetric sample, they are  a symmetric mode (SPM) and an antisymmetric plasma mode (ASPM).  They are  derived in a macroscopic approach using the Drude formula for the dielectric function which is valid in the limit of large  inelastic scattering time $\tau \to \infty$ and are  discussed in  Section II  and denoted as $ E_\pm$.
We stress that retardation is important to obtain the correct dispersion for small  $k-$vectors, $k_\parallel$ (parallel to the surfaces of the sample,  assumed to be planar). The energy scale which characterizes the plasma  excitations, which couple to the photonic field,  is the plasma frequency $ \omega _p$, or, better $ \omega _p/\sqrt{2}$ (see Fig.\ref{dispersion}).

 The  contribution due to the  continuum of TE  modes has been estimated  by a simple analogical  scattering model where the bulk material is reduced to a potential made of    two $\delta -$  repulsive functions at distance $2d$ along the $\hat {z}-$direction,   which provide elastic transmission and reflection  of the incoming  wave.  The  linear thickness of the sample $a$ has been related to a full size of $4d=a$. The model is presented in Section III.C. As the model has only qualitative relevance, we did not even include  difference in the propagation velocity between vacuum and material, for simplicity.  The model is quite useful, though, because, when continued analytically to imaginary energies, it allows to get an estimate of the total Casimir energy $E_{Tot}^n$, including the plasma modes\cite{bordagRep}.  At very small sizes $a$, the contribution given by  the continuum of states  to $E_{Tot}^n$ is expected to be minor and  we have used the information coming from  $E_{Tot}^n$, by shifting the curve of the corresponding energy vs linear size $a$ so to match $E_++E_-$ at small $a$.  It turns out that the discrepancy between   $E_{Tot}^n$ and  $E_++E_-$ only occurs for large $a$ values, in a range of $a$ values which  is not reliable for  reasons that will be explained below. The model is part of a more general model which  includes two channels to be described below, presented in Section III. Analysis of the extended model shows that, when the size $a$ increases beyond $a >1.5$  undesired resonant states  are produced  in the elastic channel (see Fig.\ref{shifts} {\sl inset}). This is the reason why  the model should not be accepted at large $a$ values.

 Modelization of the superconducting phase requires  three energy scales. The highest one  in energy  is the Anderson-Higgs threshold $ \hbar c m$ ($c$ is the propagation velocity in the medium and $m = 2\pi /\lambda _L^{\perp}$). Photons acquire the AH mass and a longitudinal mode arises,  eating  up the phase mode of the superconducting order parameter.  The intermediate  one is the superconducting plasma frequency  $\omega _{ps}$ and the lowest one is the Cooper  pair breaking threshold $2\Delta $.  They are discussed in Section II.  At energies below the AH threshold light does not propagate (radiation gap), unless it is coupled to quasiparticle (qp)  excitations. The difference with the normal phase is substantial in the energy window defined by the electronic superconducting gap $\Delta$.  However,  qp's can originate at finite temperature  from nodes in the gap  or  any type of  pair breaking process.  We do not consider the continuum of propagating photon states for energies above the $2\Delta$ threshold, because we have neglected the corresponding states in the normal phase and, except for marked changes in proximity of  $2\Delta $, which are in any case dropped,   we assume that this energy range of both spectra roughly cancels in the difference. The TM photon mode  has $B_z=0$ at the surfaces and satisfies the macroscopic London equation. This is the reason why we can keep a macroscopic picture when discussing  the transverse massive e.m. fields at the surfaces, each of  which roughly corresponds to  the e.m. TM  field of the normal phase.  Both of them couple to the  plasma excitations of the  sample in the superconducting phase. There are two plasma modes in the superconducting  phase of limited geometries, which can be derived in a hydrodynamic approach\cite{mooij}: the Mooij  and Sch\"on (MS) acoustic mode  and the Carlson-Goldman (CG) mode. The first one corresponds to the SPM of the normal phase and has a $\sqrt{k_\parallel}$ dispersion  and lies  within the superconducting gap (see Fig.\ref{mooij}). The CG mode  is  in proximity of the $2\Delta$ threshold  and involves qp's which neutralize the charge in a sort of ASPM. This mode, being neutral, does not couple with radiation and is ignored.  In addition, resonances can appear in the radiation gap,   even in the $2\Delta$ gap, which split off the AH threshold by virtual interaction with the Cooper pair condensates of the {\sl a-b} planes (see Fig.\ref{randa}).  They provide resonances which make the longitudinal massive mode  propagating in the superconducting gap. We have shown that this is possible  by setting up the scattering model  of Section IV, with an elastic channel and a closed channel.  Of the two resonances,  a symmetric and an antisymmetric one, only one is present at energy below $2\Delta$, depending   on the linear size of the sample.  The  antisymmetric one is only at low energies, when the size of the sample is   $a >  2 \lambda _L^{\perp}$ (see Fig.{\ref{randa}).
 
 With the mentioned approximations an estimate of the Casimir energy  difference   between the two phases $ \delta{\cal{E}} =  E_{Sup}-  E_{Nor}=  E^{TM}_{Sup} + E^{TE}_{Sup} - (E_++E_-) \sim -  10^{-5} \div10^{-6} eV$ is  reported for a few linear sizes of the sample in Fig.\ref{risot}  for a reference area $a^2$, where length are  in units of  $ c/\omega _p $ and is marked by the {\sl black dots} in the figure. The dependence on the linear size of the sample is $\sim 1/ a^4$,  for large sizes, as found in the measurement of the Casimir-Polder force\cite{sukenik}. The pair breaking threshold $2\Delta $ is also reported for comparison and longer samples imply that  the energy window in the superconducting gap shrinks. To achieve these estimates, quite different qualitative models have been invoked: a macroscopic model for the TM polarization, a scattering 'microscopic' model for the TE  polarization both in the form of  one channel elastic scattering and in the form of a two channel  scattering. As the models have little giustification and the correspondence between them is arbitrary,  the results cannot be considered as quantitative. They are just an indication of the physics involved, which should be checked carefully in the course of the experiment.    It is clear  that the largest contributions to the difference $ \delta{\cal{E}} $  arise from  the superconducting gap window and from the energy window across the pair breaking threshold  for the TM polarization (see Fig.\ref{ridifo}). The latter contribution has been qualitatively discussed in Appendix C, but has not been included in our estimate and requires further consideration. The reference linear size of the sample is $a\sim  c/\omega _p $ which is $\sim \lambda_L^\perp$ if $\omega _p \sim \omega_{ps}$. This is the choice that has been done to simplify  our estimates, but we stress  that it is the crucial point in the design of the experiment.  As discussed in SectionV.B,  an appropriate trade off between temperature and linear size of the sample is required.   $\omega _p \sim \omega_{ps}$ implies that the pair  electron density $n_s$ exhausts the total electron density $n$, but this only happens at very low temperatures
 $T<< T_c$. At these temperatures the effective linear scales of the normal and superconducting phase, which are dictated by the penetration depth of the photon field,  are of the same order, provided the sample is close to be an ideal metal ($\omega \tau >>1$). However the small value of $\lambda_L^\perp$ implies that the linear size of the sample should be small if the boundary surfaces of the sample are supposed to have  Casimir  interaction and a very homogeneous slab should  be syntetized, what reduces the measured weight. One can envisage a layered structure, which is also being considered by the team involved in the experiment.  
 
   \vspace{1cm}

	   {\bf Acknowledgement}
	   
	   \vspace{0.3cm} 
	   
	   We acknowledge useful discussions with Procolo Lucignano and Rosario Fazio. Work financially supported by the joint project of the Istituto Nazionale di Fisica Nucleare, the University of Sassari, the Istituto Nazionale di Geofisica e Vulcanologia, University of Cagliari and IGEA S.p.a (project SAR-GRAV, funds FSC 2014-2020); by University of Sassari with "Fondo di Ateneo per la ricerca, 2019 and 2020" and by University of Napoli, "Federico II",with project "time crystal", E69C20000400005.

   \vspace{1.0cm}

\begin{appendix}

   \section{ $t-$matrix and change in the density of states}

With $ G = G_0 + G_0 \: t \: G_0 $, we have:
\bea
\Delta \nu (\omega )  = - \frac{1}{\pi} \: \Im m \: Tr \left \{ G^R-G_0^R\right \}\nonumber\\
 = - \frac{1}{\pi} \: \Im m \: Tr \left \{ G_0^R\: t \: G_0^R \right \} 
\enea
\bea
 G_0^R = \left [ \omega + i 0^+ -H_0 \right ]^{-1}, \:\:\:\:\:  \frac{d G_0}{d\omega } = - G_0^2  \nonumber\\
 Tr \left \{ G_0\: t \: G_0 \right \} = Tr \left \{ G_0^2 \: t  \right \}= Tr \left \{ -\frac{d \: G_0}{d\omega } \: t  \right \} \nonumber
 \enea

 As $  t = V \: \sum_{n=0}^\infty \left ( G_0^R \: V \right ) ^n   $, we have:
  \bea
  Tr \left \{ -\frac{d \: G_0}{d\omega } \: t  \right \} = Tr \left \{ -\frac{d \: G_0}{d\omega }  \: V  \sum_{n=0}^\infty  \left [ G_0\: V  \right ]^{n}  \right \}\hspace{1cm}
\enea

\bea 
=  \sum_{n=1}^\infty  Tr \left \{ -\frac{d \: G_0}{d\omega }  \: V\left [ G_0\: V  \right ]^{n-1}  \right \}\nonumber\\ 
 = \sum_{n=1}^\infty  \frac{1}{n} Tr \left \{ -\frac{d}{d\omega } \left [ G_0\: V  \right ]^n  \right \}
 = \frac{d}{d\omega } Tr  \left \{  \ln\: \left [ 1 -  G_0^R\: V  \right ]  \right \} .\nonumber 
 \enea
 It follows that:
   \begin{widetext}
 \bea
- \Im m Tr  \left \{  \ln\: \left [ 1 -  G_0^R\: V  \right ]  \right \}  =  - \frac{1}{2\: i} \left [ Tr  \left \{  \ln\: \left [ 1 -  G_0^R\: V  \right ]  - \ln\: \left [ 1 -  G_0^A\: V  \right ] \right \} \right ] \nonumber\\
=\frac{1}{2\: i} \left [ Tr  \left \{  \ln\: \left [ \left (1 -  G_0^A\: V \right )\: \left ( 1 -  G_0^R\: V \right )^{-1}  \right ] \right \} \right ].\nonumber\\
 \left (1 -  G_0^A\: V \right )\: \left ( 1 -  G_0^R\: V \right )^{-1}  =  \left (1 -  G_0^A\: V \right )\: \left ( 1 +  G_0^R\: V + G_0^R\: V \:  G_0^R\: V + ... \right ) \nonumber\\
  = 1- \left ( G_0^A -G_0^R \right ) -  G_0^A\: V \:  G_0^R\: V + G_0^R\: V \:  G_0^R\: V +... = 1-   \left ( G_0^A -G_0^R \right ) \left [V+  VG_0^R\: V  +... \right ] = 1- \left ( G_0^A -G_0^R \right )  \: t \nonumber
\enea
\bea
 G_0^A -G_0^R  = 2 \: i\: \delta \left ( \omega -H_0 \right ) , \:\:\:\: S(\omega ) = 1 -  2  \pi i\: \delta \left ( \omega -H_0 \right )\: t(\omega ) \nonumber\\
 \Delta \nu (\omega )  = - \Im m \frac{d}{d\omega }  Tr  \left \{  \ln\: \left [ 1 -  G_0^R\: V  \right ]  \right \} = \frac{1}{2\pi  i } \frac{d}{d\omega } 
 Tr  \left \{ \ln \left[1 -  2  \pi i\: \delta \left ( \omega -H_0 \right )\: t(\omega )\right ]\right \} \nonumber\\
 =  \frac{1}{2\pi  i } \frac{d}{d\omega }\: Tr  \left \{ \ln S(\omega )\right \}  =  \frac{1}{2\pi  i } \frac{d}{d\omega }\: \ln\:  Det\: S(\omega ) =   \frac{1}{\pi  } \sum _j \frac{d}{d\omega }\: \delta _j(\omega ).
 \label{change}
 \enea
 as $ S(\omega ) =  diag\left [ e^{2\:i\: \delta _J(\omega )}\right ].$
\end{widetext}
 \section{Derivation of the $S-$matrix for scattering across one superconductor plane}

        Starting from   Eq.s(\ref{smat}, \ref{eqpart}) and projecting Eq.(\ref{eqL},\ref{dev})   onto our basis (we trace on the state of the superconducting condensate), we get  equations for  $ \alpha ,\beta$:
        \bea
   \langle 0,K |   \left | \psi _L(0) \right \rangle &\rightarrow & 1+ r  = 1+
(  \alpha+\beta) \cos \theta 
  \nonumber\\
   \langle 1,K |  \left | \psi _L(0) \right \rangle&\rightarrow&  s   =
( \alpha-\beta ) \sin\theta ) 
\label{contiL}
    \enea
    \bea
      \langle 0,K | \left | \psi _R(0) \right \rangle&\rightarrow & t = 1+
(  \alpha +\beta)\cos \theta
 \nonumber\\
   \langle 1,K | \left | \psi _R(0) \right \rangle&\rightarrow &  \tau =  ( \alpha-\beta)  \sin\theta
\label{contiR}
    \enea
 
 \bea
         \left | dev \right \rangle =   \left .  \frac{ d  \left | \psi _L \right \rangle }{dz} \right |_{z=0^-} -  \left .  \frac{ d  \left | \psi _R  \right \rangle}{dz} \right |_{z=0^+}
         = g\:\left [ | \Phi \rangle V+  | \pphi \rangle V^*\right ]\nonumber\\
   \langle 0,K |   \left | dev \right \rangle \rightarrow   i\: k ( 1-r-t)=  g\left [ V(1+
 \alpha )+ V^* \beta   \right ]  \cos \theta 
 \nonumber\\
   \langle 1,K |   \left | dev \right \rangle  \rightarrow \kappa_L s  + \kappa _R \tau  =  g\left [ V \alpha -V^* \beta   \right ]\sin \theta
\nonumber\\
\label{salto}
    \enea
    Here we observe that  
      the structure reflects the usual $\delta -$function potential in  a $1-$dimensional   Schr\"odinger equation. Continuity of wavefunction and jump in the derivative provide ( $g>0 \to $repulsive $\delta -$barrier in the following):   
\bea
 1+r = t ,\hspace{1.5cm} ik ( 1-r ) - ik t = -gt \nonumber\\
 1+r =t ,\hspace{1.5cm} 1-r  =  \left ( 1 - \frac{ g}{ i\: k} \right )  t \nonumber\\
t = \frac{1}{ \left ( 1 + \frac{i\: g}{ 2k} \right ) }  ,\hspace{0.5cm}  r= - i \: \frac{ g}{ 2k} \:t ,\hspace{0.5cm} \:\:  |t|^2 +|r|^2 =1,\hspace{0.5cm} 
\label{deltap}
\enea
We use Eq.s(\ref{salto}) together with   Eq.s(\ref{contiL}, \ref{contiR})  to derive the dependence of $\alpha,\beta $ on $ \kappa_L,\kappa_R, k$ ($\kappa_+ = \kappa_L+\kappa_R$). :
\bea
r= ( \alpha +\beta)\cos \theta, \:\: \:\:\: s=  \tau=  (\alpha-\beta) \sin \theta,    \nonumber \\
1-t = - ( \alpha +\beta)\cos \theta, \hspace{0.5cm}
 \label{reso} \\
 - 2 i\: k\: ( \alpha +\beta) =  g\left [ V(1+
 \alpha  )-V^* \beta  \right ],\nonumber\\
  \kappa_+ ( \alpha -\beta )   =  g\left [ V \alpha +V^* \beta  \right ]
\label{esta}
\enea
Solving  Eq.s(\ref{esta})  with respect to $\alpha,\beta$, we get,  to lowest order in $g$: 
  \bea
 {\alpha}  +{\beta} =  \frac{ g\: V }{2  {\kappa}_+ }\left [1 + \frac{ {\kappa}_+}{2 \: i\: k }\right ] , \:\:\:   {\alpha}  -{\beta} =   \frac{ g\: V }{2  {\kappa}_+ }\left [1 - \frac{  {\kappa}_+}{2 \: i\: k }\right ].\hspace{0.4cm} 
    \label{alpbet}
   \enea
   
   
   $\kappa $ should depend on  the interaction $\hat V$, but,  in the absence of information about   the interaction $\hat V$, we take  it as a function of the $\kappa $'s themselves. 
 We take 
  \bea
  \frac{gV}{2\kappa _+}=  -i\:  \frac{ {\kappa}_+}{2 \: k } \: \frac{1}{1+  \left (\frac{ {\kappa}_+}{2 \: | k  | } \right )^2}.
  \label{pot}
  \enea
  This choice is consistent wth unitarity of the  $S-$matrix which  implies $S^\dagger= S^{-1}$:
    \begin{widetext}
    \bea 
S^\dagger =  \left (  \begin{array}{cc} r ^\dagger & {t }^\dagger \\{t'}^\dagger & {r'}^\dagger \end{array}\right ) =
 \left (  \begin{array}{cc}\left [ r-t'  \: {r'}^{-1} t\right ]^{-1} & -\left [ r-t '\:  {r'}^{-1} t\right ]^{-1} t' \: {r'}^{-1} \\
 -{r'}^{-1}t\left [ r-t'  \: {r'}^{-1} t\right ]^{-1} & \left [ r'-t  \: {r}^{-1} t'\right ]^{-1} \end{array}\right ) 
   \label{smat1}\\
   {r^\dagger }^{-1} = r - t '\: {r'}^{-1} t, \hspace{0.5cm}    {t'}^{\dagger -1} = t'  - r \: {t}^{-1} r'
   \label{rela}
    \enea  
    \end{widetext}
 Consistency of Eq.(\ref{pot}) can be easily seen in the case of a single channel with  $t=t'$ for time reversal invariance.  From unitarity:
   \bea
  \frac{r}{{r'}^*} = - \frac{t}{t^*},
  \label{cruc}
  \enea
  so that, if we substitute this into Eq.s(\ref{rela}) we get:
  \bea
 r-   {r^* }^{-1} = - {r^*}^{-1} t\: t^*, \hspace{0.5cm}    {r^* }^{-1} =    r \:  \left [1- t {t^*}\right ]^{-1}\label{relar}\\
    {t^* }^{-1} -t =  {r^*} \:r\:  t^{-1} \: t  {t^* }^{-1} , \hspace{0.5cm}    \left [1-   {r^*} \: r \right ] {t^* }^{-1} =  t\label{relar1}
    \enea  
    As $ r^*r + t\: t^*  =1$,  the second of Eq.(\ref{relar1}) is $t=t$, while the second of  Eq.(\ref{relar}) is satisfied by 
    $r = t-1 $ if $ t=  1/ (1+ i  \frac{ {\kappa}_+}{2 \: k } )$.   The result is,
    \bea
     t =\frac{1}{1+ \frac{i\: \kappa _+ }{2k}}, \hspace{0.5cm}    r= t-1 =  - \frac{i\: \kappa _+}{2k}  \: t ,
     \enea
     which is what is found in case of a $\delta-$function potential. 
     \section{Signature of the pair breaking processes in the dielectric function at  $ k_0 = \frac{2\Delta }{\hbar c } $} 
          An approximate  comparison between $\epsilon_1(\omega )$ for the normal and the superconducting phase is reported  in Fig.\ref{ridifo}. According to Eq.(\ref{glov}) and the arguments given above, the two functions should acquire the same functional behavior at very low temperature, both at low and high frequencies, if  the metal is assumed close to being ideal.  
     
 At very low temperature   and frequency  the difference    is very small, due to the contribution  of the $\delta-$function at zero frequency  to the  Kramers Kroenig transform of Eq.(\ref{opi}), with $ \omega _{ps} \approx \omega _p$. In fact, the  $\delta-$  zero frequency peak  of the superconducting phase  provides $\epsilon (\omega )$ given by  Eq.(\ref{glov}), which is the same as in the case of an ideal normal metal (with $\omega \tau >>1$). Increasing the temperature  the quasiparticles contributing to  the normal metal phase are absent in the superconductor, inside the energy gap and a difference emerges. 
 In Fig.\ref{ridifo} we report the difference   between the superconducting and normal metal response at microwave  frequency,  which vanishes at zero temperature well below  $2\Delta/\hbar$. 
 The sharp peak at $ \omega \sim 2\Delta /\hbar$ heralds the enhancement of qp excitations at the pair breaking energy. Correspondingly, there is a dip in the in the mode dispersion of the  superconducting phase, as compared to the normal phase, which is concentrated at the pair-breaking frequency $\omega \tau =2$. This can be seen by  comparing the two equations derived  from Eq.(\ref{quat2})  for the symmetric mode, between  the normal and superconducting case. 
\bea
- \epsilon_S(\omega_S) =  \frac{ \kappa_S }{\kappa_0}\: \coth\frac{ \kappa_S a }{2}, \:\: - \epsilon_N(\omega_N) =  \frac{ \kappa_N }{\kappa_0}\: \coth\frac{ \kappa_N a }{2}. \nonumber
\enea
 If we neglect retardation in this frequency range, $\kappa_S \approx  \kappa_N\approx \kappa_0= k_\parallel$, so that, with $ \epsilon_S(\omega_S)\approx  \epsilon_S(\omega_N) + \delta \omega \: \left . \frac{ \partial  \epsilon_S(\omega)}{\partial \omega } \right |_{\omega _N} $, we obseve that 
 \beq
  \epsilon_S(\omega_N)  -\epsilon_N(\omega_N)+ \delta \omega \: \left . \frac{ \partial  \epsilon_S(\omega)}{\partial \omega } \right |_{\omega _N}  =0.
  \eneq
  Immediately before the peak the difference is positive and the derivative is positive, so that $\delta\omega <0$. Immediately after the peak the derivative becomes negative so that $\delta\omega <0$ and they form a cusp pointing downward. After the peak the difference is negative and the derivative is positive, so that $\delta\omega >0$ increases again.   The location of the cusp is about $ k_0 = \frac{2\Delta }{\hbar v } $ where $v \sim 10^8 cm/sec$  is the velocity of the electron in the metal, giving $k_0= 250 \: (\mu m)^{-1}$   which is  a $k-$vector  sampling distances of the order of the lattice spacing, beyond the validity of this approach.

     \end{appendix}

\end{document}